\documentclass[11pt,a4paper]{article}
\pdfoutput=1 

\usepackage[table]{xcolor}
\usepackage{colortbl}
\RequirePackage{ifpdf} 
\usepackage{amsmath} 
\usepackage{mathtools}
\usepackage{jheppub}
\usepackage{pstricks}
\usepackage[final]{pdfpages} 
\usepackage{ifpdf} 
\usepackage{slashed}
\usepackage{float}
\usepackage{comment}
\usepackage[caption = false]{subfig}
\usepackage[normalem]{ulem}
\usepackage{color}
\usepackage{xcolor}
\definecolor{urlblue}{rgb}{0.2,0.4,0.7}
\definecolor{citegreen}{rgb}{0,0.6,0.2}
\definecolor{linkred}{rgb}{0.9,0.2,0.1}
\usepackage{hyperref}
\hypersetup{
colorlinks=true, citecolor=citegreen, linkcolor=blue, urlcolor=urlblue}

\usepackage{graphics}
\usepackage{etoolbox} 
\usepackage{fixmath}
\usepackage{psfrag}
\usepackage{textgreek}

\usepackage{notoccite} 

\usepackage{amsfonts}
\usepackage{autobreak}
\usepackage{marginnote}
\usepackage{enumitem}
\usepackage{appendix}
\usepackage{caption} 
\newcommand{\NOdisplay}[1]{ }

\title{QCD corrections to the Golden decay channel of the Higgs boson}

\author{Mandeep Kaur$^{a}$, Maguni Mahakhud$^{a,b}$, Ambresh Shivaji$^{a}$, Xiaoran Zhao$^{c}$}
\emailAdd{mandeepkaur.iiser@gmail.com, maguni.mahakhud@kolhanuniversity.ac.in, ashivaji@iisermohali.ac.in, 
xiaoran.zhao@uniroma3.it}
\affiliation{
$^a$Indian Institute of Science Education and Research Mohali,\\
Knowledge City, Sector 81, SAS Nagar, Manauli, Punjab 140306, India
\\$^b$University Department of Physics, Kolhan University,\\
Chaibasa, West Singhbhum, Jharkhand- 833201
\\$^c$Dipartimento di Matematica e Fisica, Universit\`{a} di Roma Tre and INFN,\\sezione di Roma
Tre, I-00146 Rome, Italy}

\preprint{}

\abstract{{Future colliders aim to provide highly precise experimental measurements of the properties of the Higgs boson. In order to benefit from these precision machines, theoretical errors in the Higgs sector observables have to match at least the experimental uncertainties. The theoretical uncertainties in the Higgs sector observables can be reduced by including missing higher-order terms in {{their}} perturbative calculations.} In this direction, we compute the mixed QCD-electroweak corrections at ${\mathcal O}(\alpha \alpha_s)$ to the Higgs decay into four charged leptons by considering the golden decay channel, $ H \to e^+e^-\mu^+\mu^-$. 
 Due to color conservation, these corrections receive contributions only from the two-loop virtual diagrams. In the complex mass scheme, 
we find that the mixed QCD-electroweak corrections to the partial decay width, relative to the leading order predictions, are positive and about $0.27\%$ for $\alpha_s(M_Z)$. Relative to the next-to-leading order electroweak corrections, the mixed QCD-electroweak corrections are found to be approximately 
 $18\%$ for $\alpha_s(M_Z)$. {With respect to the leading order, we observe a flat effect of the mixed QCD-electroweak corrections on the invariant mass distribution of the lepton pairs with fixed QCD coupling. The $\phi$ distribution, due to the mixed QCD-electroweak corrections, follows a $(1-\cos \phi)$ dependence.  }
}

\begin{document}
\allowdisplaybreaks[4]
\unitlength1cm
\maketitle
\flushbottom

\section{Introduction}
\label{sec:intro}
With the groundbreaking discovery of the Higgs boson by the CMS and ATLAS~\cite{CMS:2012qbp,ATLAS:2012yve} collaborations at the LHC in 2012, particle physics has entered the realm of precision studies. The Higgs boson is one of its kind in the Standard Model (SM); therefore, precision measurements in the Higgs sector provide an opportunity to look for the physics beyond the Standard Model (BSM). {All the present and future colliders, such as the Large Hadron Collider (LHC)~\cite{Peskin:2013xra}, High-Luminosity LHC (HL-LHC)~\cite{Apollinari:2015wtw}, Future Circular Collider (FCC-ee)~\cite{TLEPDesignStudyWorkingGroup:2013myl}, Circular Electron Positron Collider (CEPC)~\cite{CEPCStudyGroup:2018rmc,CEPCStudyGroup:2018ghi}, and the International Linear Collider (ILC)~\cite{Baer:2013cma} aim to explore the uncharted territory of the fundamental interactions by measuring various Higgs properties with higher statistics. On the theory side, highly precise predictions for the Higgs production and decay channels are needed for a fairer comparison with future experimental data.}

Among the five prominent decay modes of the Higgs, a rare but important one is the decay of the Higgs boson into four charged leptons, also known as the ``Golden decay channel". This decay channel played a significant role during the Higgs discovery in 2012 as it provided a particularly clean signature around the Higgs mass ($\sim 125 $ GeV) in the invariant mass spectrum of the final state leptons. Furthermore, kinematic distributions of the final state leptons for this decay mode not only allow for precision mass measurements of the Higgs boson but also serve as a powerful tool to study its spin and $CP$ properties~\cite{ATLAS:2014euz,CMS:2013fjq,Bolognesi:2012mm,CMS:2014nkk,ATLAS:2013xga}. The data collected in the off-shell production and decay of the Higgs to four leptons via the $Z$-boson pairs can constrain its total decay width~\cite{Caola:2013yja,ATLAS:2015cuo,CMS:2014quz}. 
Thus, improved predictions for the golden decay mode are paramount to having a better understanding of the Higgs properties. In this direction, several works have been reported in the literature.

The exact one-loop QED corrections of $\mathcal{O}(\alpha )$ to the Higgs decay into four leptons with the off-shell $Z$-bosons have been evaluated in~\cite{Piccinini:2005iu,CarloniCalame:2006vr}. The complete one-loop electroweak corrections for the leptonic, semi-leptonic, and hadronic final states, and the one-loop QCD corrections for the semi-leptonic and hadronic final states to the decay $H \to WW/ZZ \to 4f$ have already been evaluated in~\cite{Bredenstein:2006rh,Bredenstein:2006ha} and are encoded in a Monte Carlo (MC) code {\tt Prophecy4f}~\cite{Bredenstein:2006nk,Bredenstein:2007ec}. The $\mathcal{O}(\alpha)$ corrections, reported for the case of four charged leptons in the final state, are of the order of 2-4\% for moderate Higgs masses ($M_H \leq 200$ GeV) and increase with the growing Higgs mass, reaching up to 14\%. In addition to that, the one-loop electroweak and QCD corrections to the Higgs decay into four fermions in the context of a simple extension of the SM have also been studied and are implemented in the code {\tt Prophecy4f}~\cite{Altenkamp:2017ldc,Altenkamp:2017kxk,Altenkamp:2018bcs}. The next-to-leading order (NLO) electroweak corrections to the Higgs decay into charged leptonic final states $H \to Z^{(*)} Z^{(*)} \to 4\ell$ with $4\ell = 4e, \;4\mu,\; 2e2\mu$ matched with QED Parton Shower (PS), have also been calculated, for which the results are available in a public event generator, {\tt Hto4l}~\cite{Boselli:2015aha}. 

In the present work, we compute the QCD corrections to the decay $H \to e^+e^-\mu^+\mu^-$ on top of the electroweak corrections that mainly receive contributions from the two-loop diagrams, appearing at $\mathcal{O}(\alpha \alpha_s)$ in the perturbation theory. These mixed QCD-electroweak corrections are expected to be small because of the two-loop effect, as compared to the NLO electroweak corrections. However, these are essential to provide precise predictions for the Higgs sector observables at the LHC and future colliders and {{to}} test the validity of the perturbative QFT calculations. Our motive is to quantify these corrections, simulate decay events, and provide improved numerical predictions for the partial decay width of $H\to e^+e^-\mu^+\mu^-$, and relevant kinematic distributions with an accuracy of $\mathcal{O}(\alpha \alpha_s)$. {The two-loop diagrams contributing to the amplitude at $\mathcal{O}(\alpha \alpha_s)$ are very similar to those appearing in the processes $e^+ e^- \to Z H$ and $e^+ e^- \to \mu^+ \mu^- H$, for which the mixed QCD-electroweak corrections of $\mathcal{O}(\alpha \alpha_s)$ have been evaluated in~\cite{Sun:2016bel,Gong:2016jys,Chen:2018xau}}. In this work, numerical calculation of the two-loop amplitude is performed systematically using our in-house codes, and finally, to provide improved predictions for the partial decay width, phase-space integration over the final-state leptons is performed by interfacing our codes with the publicly available code {\tt{Hto4l}}~\cite{Boselli:2015aha}.

The rest of the paper is organized as follows: In section~\ref{sec:QCD-cor}, we classify the Feynman diagrams that contribute to $\mathcal{O}(\alpha \alpha_s)$.
The organization of the matrix elements in terms of form-factors and their divergence structure is discussed in section~\ref{sec:FF}. In section~\ref{sec:renorm}, the UV renormalization of the two-loop matrix elements, along with the opted renormalization scheme, is described. The numerical implementation of the two-loop matrix elements for event generation and the checks performed on them are given in section~\ref{sec:imple} followed by our numerical results in section~\ref{sec:results}. Finally, we draw conclusions from our work in section~\ref{sec:conclusions}. 

\section{QCD correction to $H \to e^+ e^- \mu ^+ \mu ^-$}
\label{sec:QCD-cor}
\begin{figure}[!h]
\begin{center}
\includegraphics[width=4.2cm,height =4.2cm]{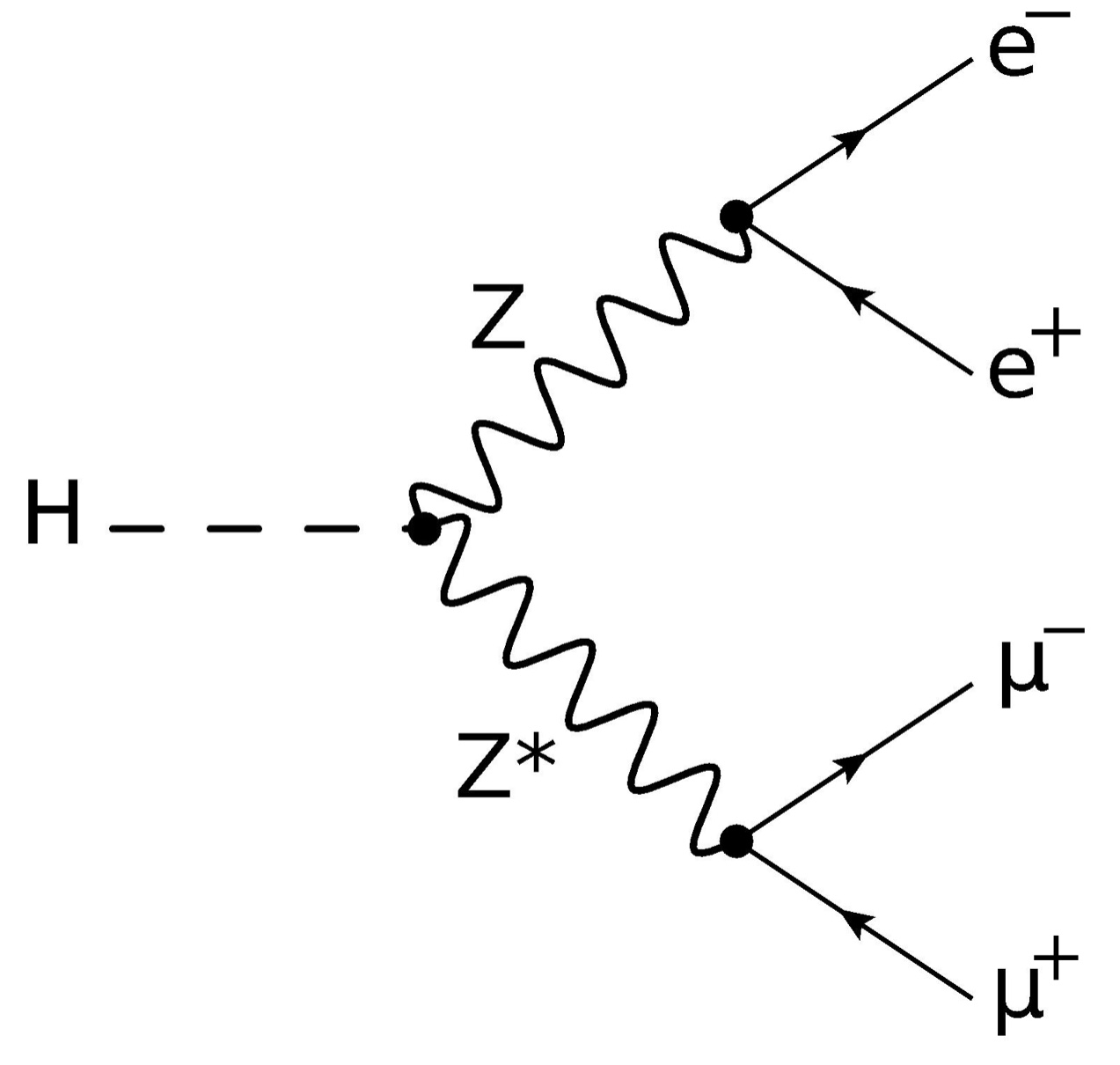}
\end{center}
\caption{LO Feynman diagram contributing to $H \to e^+e^-\mu^+\mu^-$.}
\label{fig:LO}
\end{figure}

{In the SM, the leading order (LO) contribution to the on-shell decay of the Higgs boson into four charged leptons ($ H \to e^+ e^- \mu^+ \mu^-$), mediated by a pair of $Z$-bosons, comes from a tree-level diagram, shown in Fig.~\ref{fig:LO}. Due to energy conservation, at least one of the two $Z$-bosons has to be off-shell, depicted by $Z^*$. In this work, we consider a more general case by treating both the $Z$-bosons off-shell.} We choose the following momenta assignments for the particles in the decay.  
\begin{equation}
H(q) \rightarrow Z^{(*)}(p_1)Z^{(*)} (p_2) \rightarrow e^+(p_3) e^-(p_4) \mu^+(p_5) \mu^-(p_6),
\end{equation}
where momentum conservation requires $q =( p_1 +p_2)$, $p_1 = (p_3 + p_4)$ and $p_2 = (p_5 + p_6)$. In our calculation, we are neglecting the masses of the final-state leptons. Various scalar products of interest are given by, 
\begin{equation}
p_3^2 = p_4^2=p_5^2=p_6^2=0, \; \;p_1^2 = 2p_3.p_4,\; \;p_2^2 = 2 p_5.p_6,\; \; q^2 = m_H^2,
\end{equation}
where $m_H$ is the Higgs mass. As the decay width is proportional to the squared amplitude, the amplitude for $H \to Z^{(*)} Z^{(*)} \to e^+ e^- \mu^+ \mu^-$ in the perturbative expansion up to two-loop order can be written as,
\begin{equation}
M_{\text{Total}} = M_0 + M_1^{\alpha} + M_2^{\alpha \alpha_s}+ \dots
\end{equation}
Here $M_0$, $M_1$, and $M_2$ are the LO, one-loop, and two-loop amplitudes. {In the SM, $M_1$ receives contribution only 
from the electroweak (EW) sector, as the particles involved at the LO are color neutral. However, at the two-loop level, both the EW and QCD sectors can contribute. Therefore, there can be contributions of $\mathcal{O}(\alpha^2)$ and $\mathcal{O}(\alpha \alpha_s)$ at the two-loop level, but one can neglect the contributions of $\mathcal{O}(\alpha^2)$ due to the smallness of the EW coupling $\alpha$ in comparison to the strong coupling $\alpha_s$. Thus, we only consider the mixed QCD-electroweak corrections of $\mathcal{O}(\alpha \alpha_s)$  at the two-loop level and focus on the evaluation of 
$M_2^{\alpha \alpha_s}$.

{In going beyond the LO, $\mathcal{O}(\alpha)$ amplitude receives contributions from several one-loop diagrams mediated by the weak bosons and quarks. However, only quarks are susceptible to couple with the gluons and take part in the QCD corrections; therefore, at $\mathcal{O}(\alpha \alpha_s)$, only the diagrams with quark loop along with the gluon dressing will contribute. The contribution to the amplitude at $\mathcal{O}(\alpha \alpha_s)$ can be divided into three categories as follows:
\begin{equation}
\label{eqn:Total_Amp}
 M_2^{\alpha \alpha_s} = \delta M_{HV_1V_2}^{\alpha \alpha_s} + \delta M_{S.E.}^{\alpha \alpha_s} + \delta M_{Z\ell \bar{ \ell}}^{\alpha \alpha_s}\;,
\end{equation}
where $\delta M_{HV_1V_2}^{\alpha \alpha_s}$ consists of corrections coming from the $HV_1V_2$ vertex, $\delta M_{S.E.}^{\alpha \alpha_s}$ contains corrections due to the self-energy insertions on the vector-boson legs, and $\delta M_{Z\ell \bar{ \ell}}^{\alpha \alpha_s}$ appears due to $\mathcal{O}(\alpha \alpha_s)$ counter-term for the $Z\ell \bar{ \ell}$ vertex, respectively. These contributions are described below.}

\subsection{$HV_1V_2$ vertex corrections}
At the two-loop level, in addition to the decay of the Higgs into $e^+ e^- \mu^+ \mu^-$ through the $Z^{(*)}Z^{(*)}$ channel, we also have to consider the contributions coming from the $Z^{(*)}\gamma^{(*)}$, and $\gamma^{(*)} \gamma^{(*)}$ channels. Thus, we consider the most general amplitude for the $HV_1V_2$ vertex corrections denoted by $\delta M_{HV_1V_2}^{\alpha \alpha_s} (V_1,V_2 = Z,\gamma)$, where 
both $V_1$ and $V_2$ are taken off-shell, and write,
\begin{equation}
\delta M_{HV_1V_2}^{\alpha \alpha_s} = \delta M_{HZZ}^{\alpha \alpha_s} +\delta M_{HZ\gamma}^{\alpha \alpha_s} +\delta M_{H\gamma Z}^{\alpha \alpha_s} + \delta M_{H\gamma \gamma}^{\alpha \alpha_s} .
\end{equation}
In the above, the $HZ\gamma$ and $H\gamma Z$ contributions are written explicitly to take care of the fact that $e^+e^-$ can come from either $Z$ or $\gamma$.
{{As}} the leptonic decay of the vector bosons is not affected by the QCD corrections to the $HV_1V_2$ vertex, we can decompose the amplitude for the $HV_1V_2$ vertex corrections as
\begin{equation}
\delta M_{HV_1V_2} = M^{\mu \nu} J_{\mu}(p_1)J_{\nu}(p_2),
\end{equation}
where, $M^{\mu\nu}$ is the two-loop amplitude for $H (q) \rightarrow V_1(p_1)V_2(p_2)$ decay; $J_{\mu}(p_1)$ and $J_{\nu}(p_2)$ are the fermionic currents corresponding to $V_1 \rightarrow e^+ e^-$ and $V_2 \rightarrow \mu^+ \mu^-$ respectively. {Since coupling of the top quark with the Higgs is the largest among all the quark flavors, we neglect the contributions from diagrams with any quark other than the top quark in the loop.} There are in total 48 two-loop triangle diagrams contributing to the $H (q) \rightarrow V_1(p_1)V_2(p_2)$ decay at $\mathcal{O}(\alpha \alpha_s)$, out of which some representative diagrams are shown in Fig.~\ref{fig:triangle}. 
\begin{figure}[ht!]
\centering
\subfloat[]{\includegraphics[width = 1.4in]{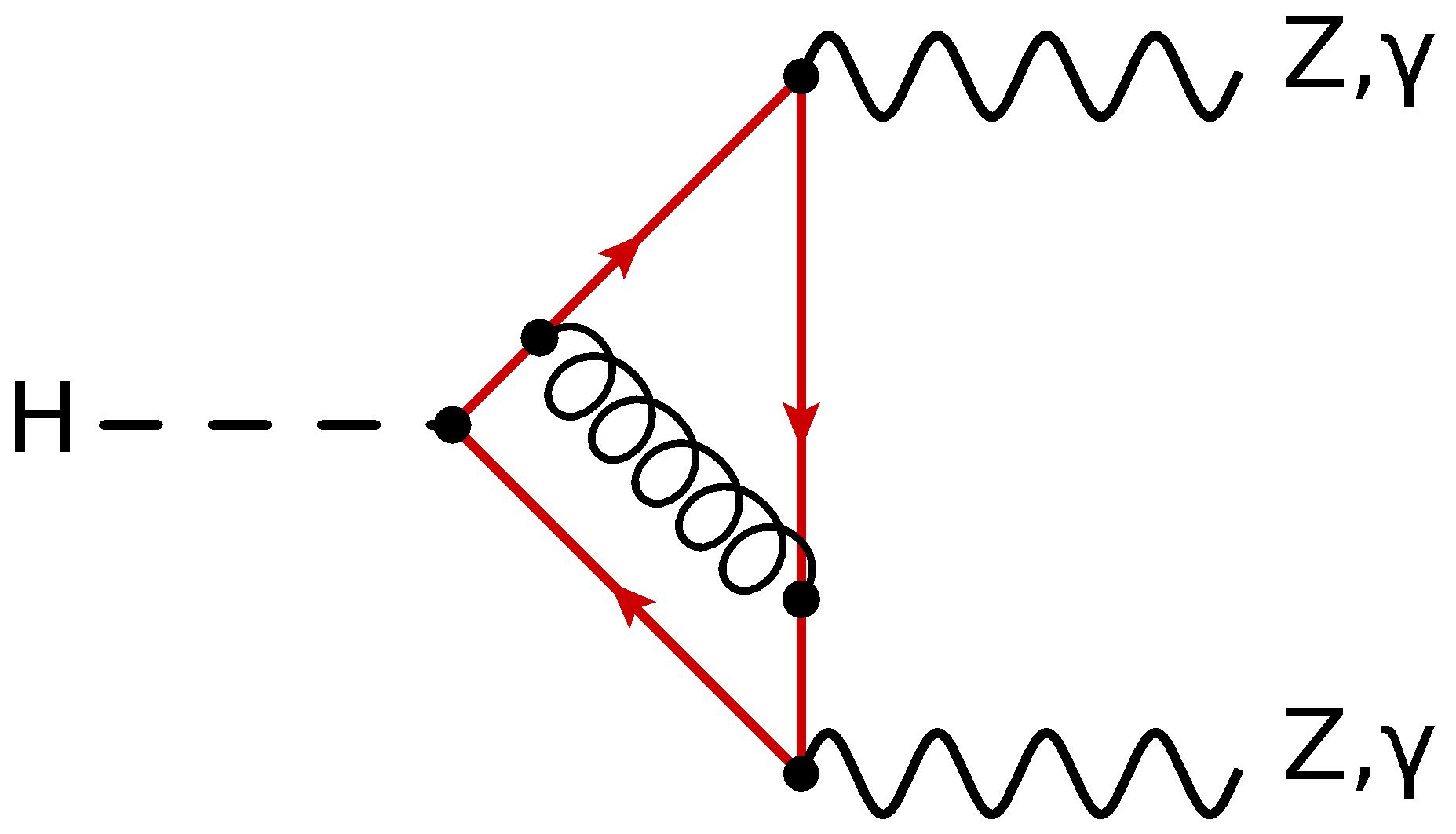}} 
\hspace{1cm}
\subfloat[]{\includegraphics[width = 1.4in]{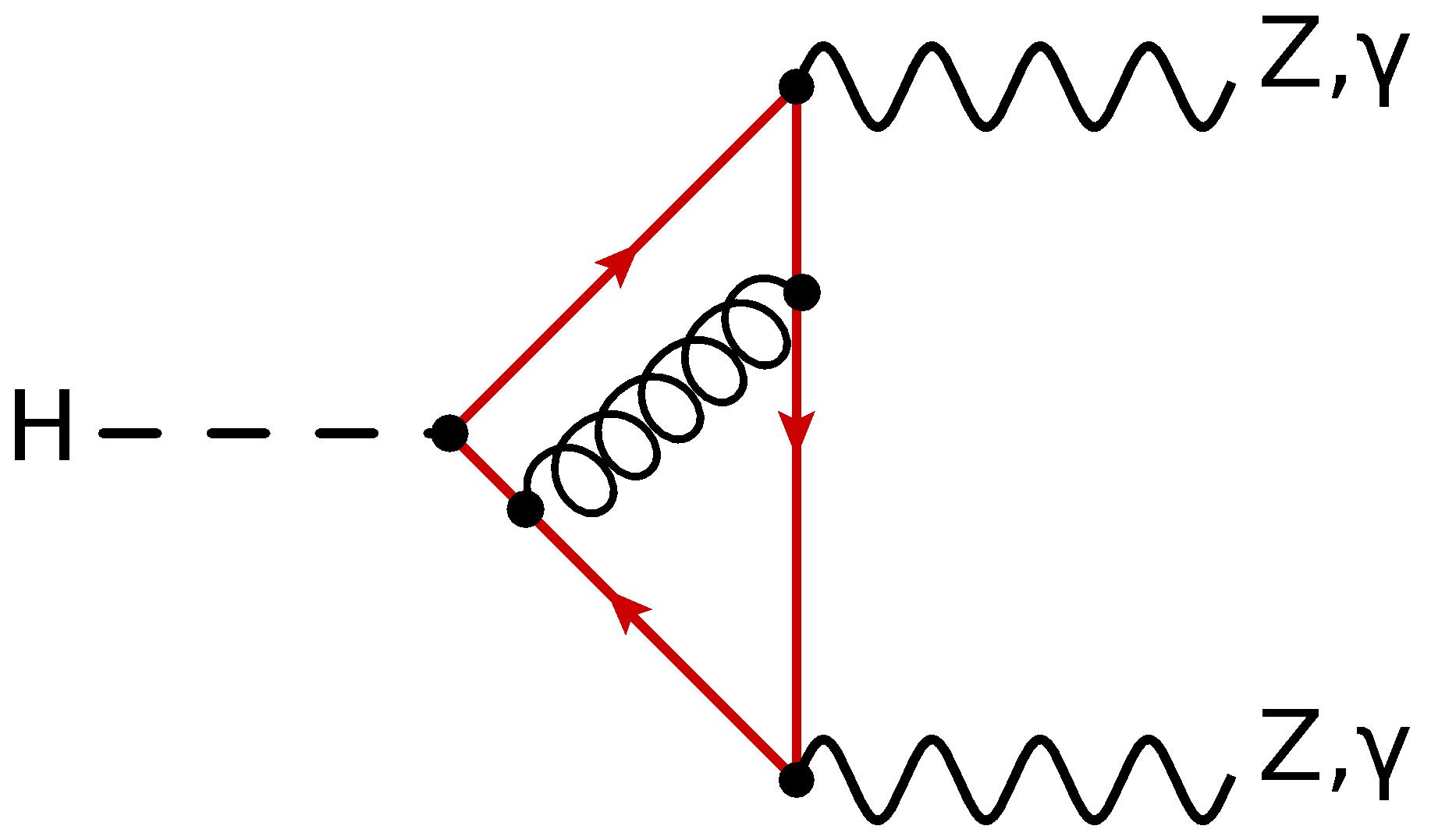}}
\hspace{1cm}
\subfloat[]{\includegraphics[width = 1.4in]{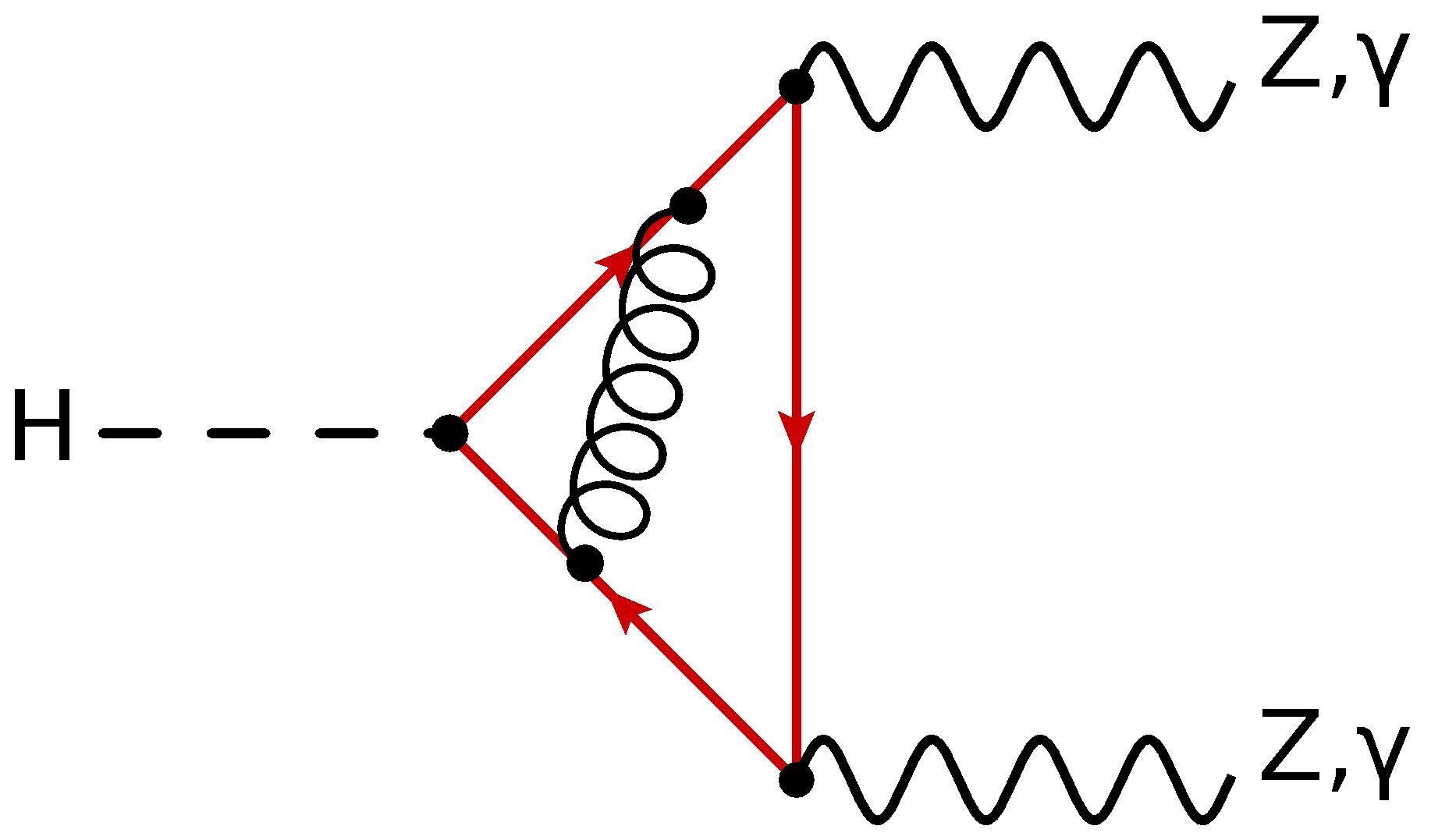}}\\ 
\subfloat[]{\includegraphics[width = 1.4in]{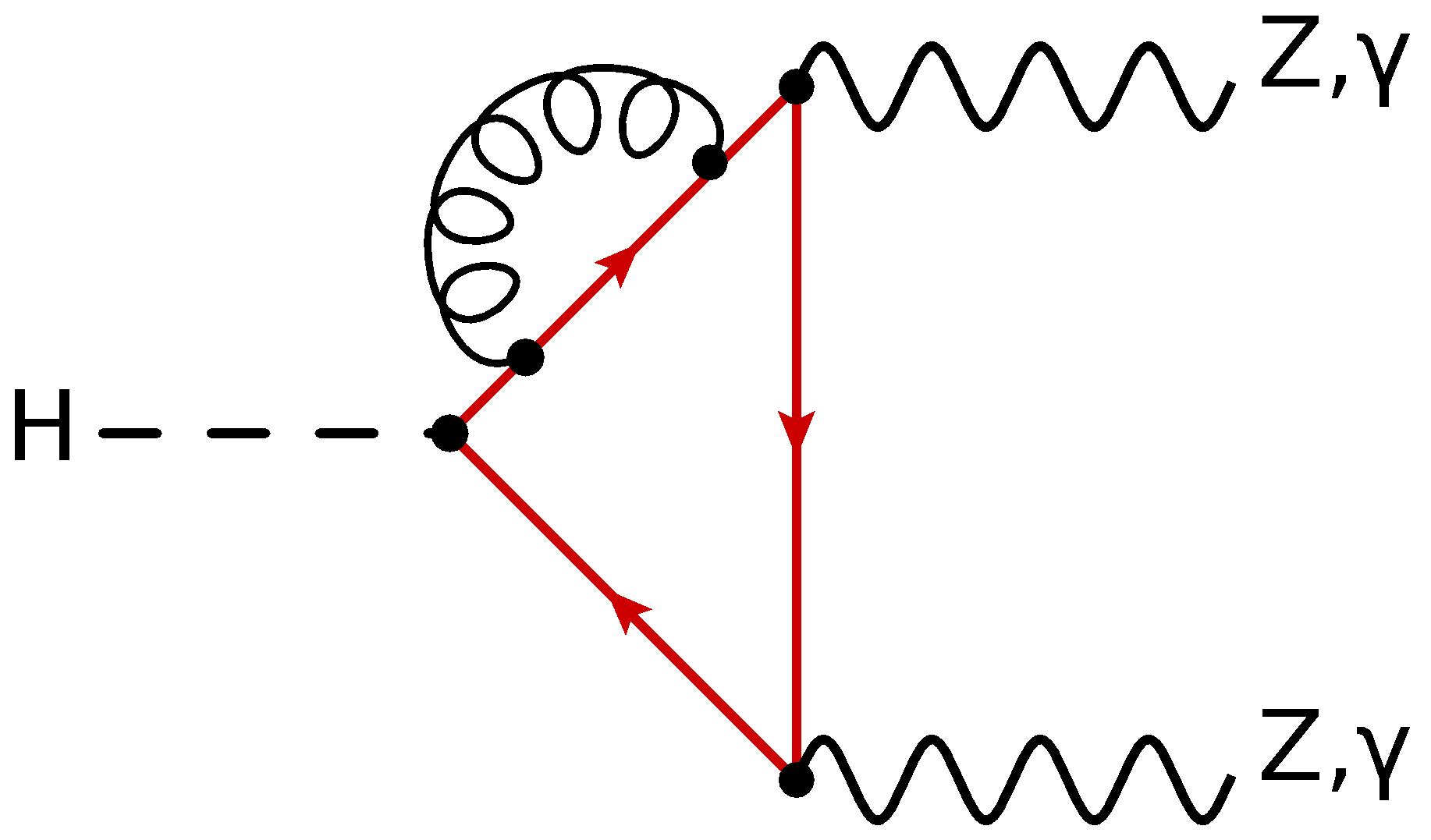}} 
\hspace{1cm}
\subfloat[]{\includegraphics[width = 1.4in]{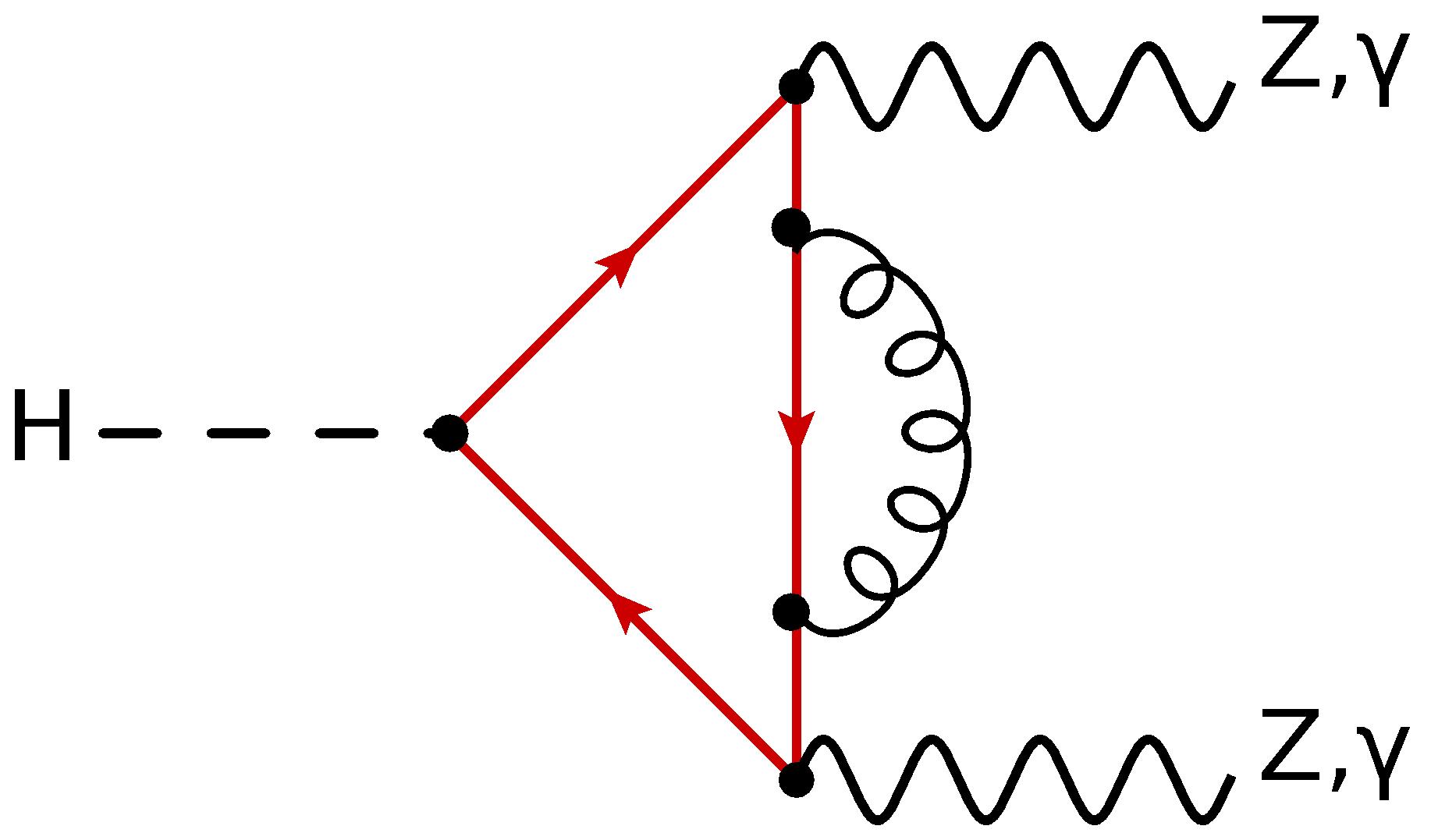}}
\hspace{1cm}
\subfloat[]{\includegraphics[width = 1.4in]{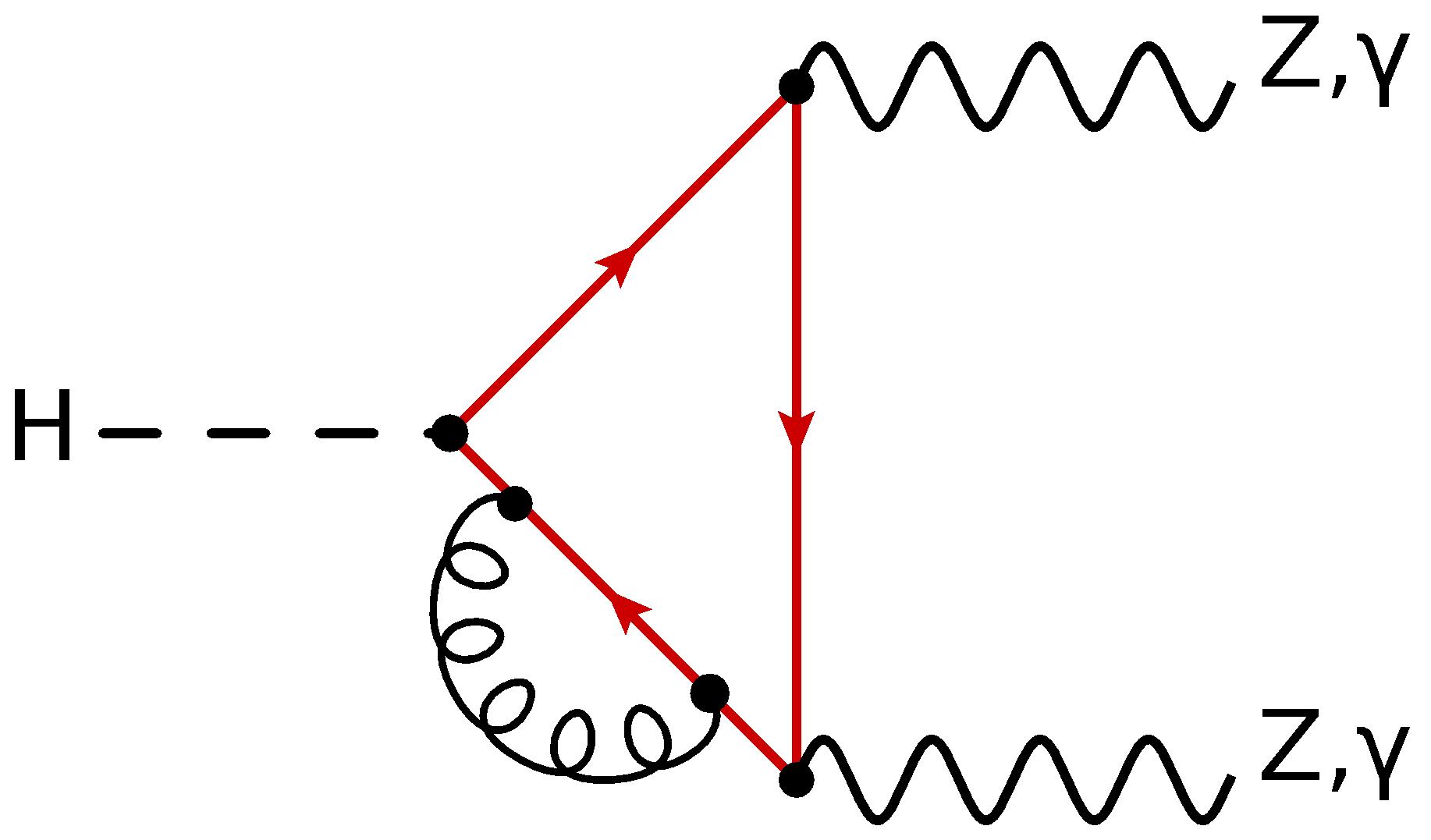}} \\
\caption{Representative two-loop triangle diagrams contributing to the bare amplitude of $H \to V_1 V_2 \;(V_1,V_2 = Z,\gamma)$ with the top quark running in the loop. Diagrams with the reversed direction of the fermionic current are not shown.}
\label{fig:triangle}
\end{figure}
The diagrams are generated with the help of the \texttt{QGRAF}~\cite{NOGUEIRA1993279} package. The amplitude for each of these diagrams is organized in \texttt{FORM} \cite{Vermaseren:2000nd,Vermaseren:2008kw} and is manipulated using \texttt{Mathematica}.

Instead of using the conventional methods, the projector technique has opted to perform the amplitude evaluation more systematically. In this technique, the amplitude for $H \to V_1 V_2$ using the Lorentz covariance can be expressed as
\begin{equation}
\label{eqn:general amplitude}
 M^{\mu\nu} = (A\; g^{\mu\nu} + B\; p_1^{\nu} p_2^{\mu} + C\; \epsilon^{\mu\nu p_1p_2} + D\; p_1^{\nu} p_1^{\mu} + E\; p_2^{\nu} p_2^{\mu} + F\; p_1^{\mu} p_2^{\nu}),
\end{equation}
where $A$, $B$, $C$, $D$, $E$ and $F$ are scalar functions called form-factors and $\epsilon^{\mu\nu p_1p_2} = \epsilon^{\mu\nu \rho\sigma}p_{1\rho}p_{2\sigma}$. The form-factors can be obtained by applying suitable projectors $P^i_{\mu\nu}$ ($i=A,B,C,D,E,F$) on the amplitude $M^{\mu\nu}$. These form-factors are, in general, functions of the Mandelstam variables present in the problem under consideration. The two-loop form-factors contributing to the amplitude at $\mathcal{O}(\alpha \alpha_s)$ are discussed in section~\ref{sec:FF}.

\subsection{Self-energy corrections}
The self-energy part of the two-loop amplitude, denoted by $\delta M_{S.E.}^{\alpha \alpha_s}$, receives contributions from $\mathcal{O}(\alpha\alpha_s)$ corrections to the $ZZ$ and mixed $Z\gamma$ self-energies. This part of the amplitude receives contributions from the top quark as well as from the light quarks. In total, 72 self-energy diagrams contribute to the $\delta M_{S.E.}^{\alpha \alpha_s}$, out of which some are shown in Fig.~\ref{fig:SE_bare}. In order to calculate $\delta M_{S.E.}^{\alpha \alpha_s}$, we need $\mathcal{O}{(\alpha\alpha_s)}$ expressions of the gauge boson self-energies. Their analytical expressions are available in \cite{Djouadi:1993ss,Dittmaier:2015rxo}.
\begin{figure}[ht!]
\centering
\subfloat[]{\includegraphics[width = 1.4in]{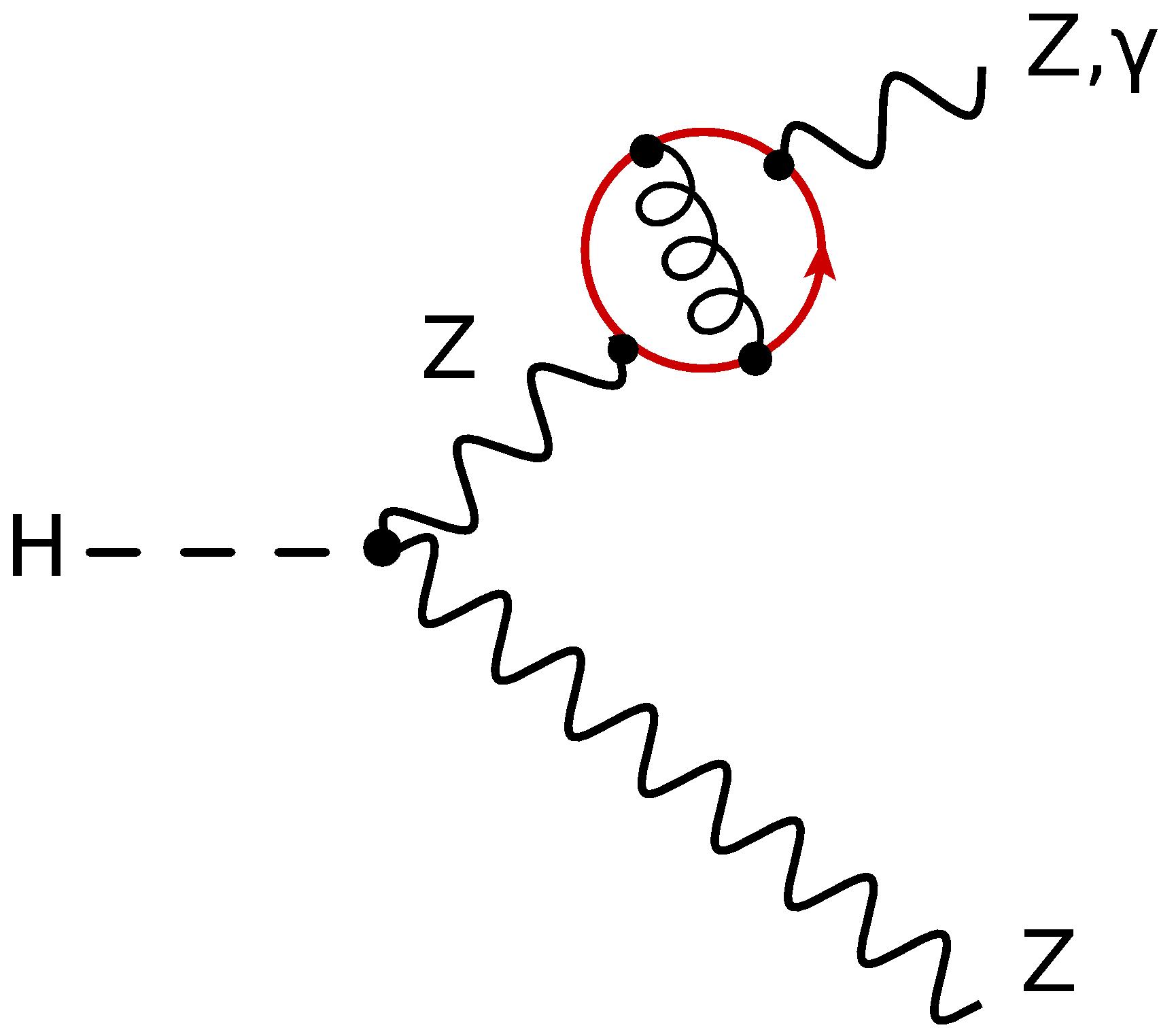}} 
\hspace{1cm}
\subfloat[]{\includegraphics[width = 1.4in]{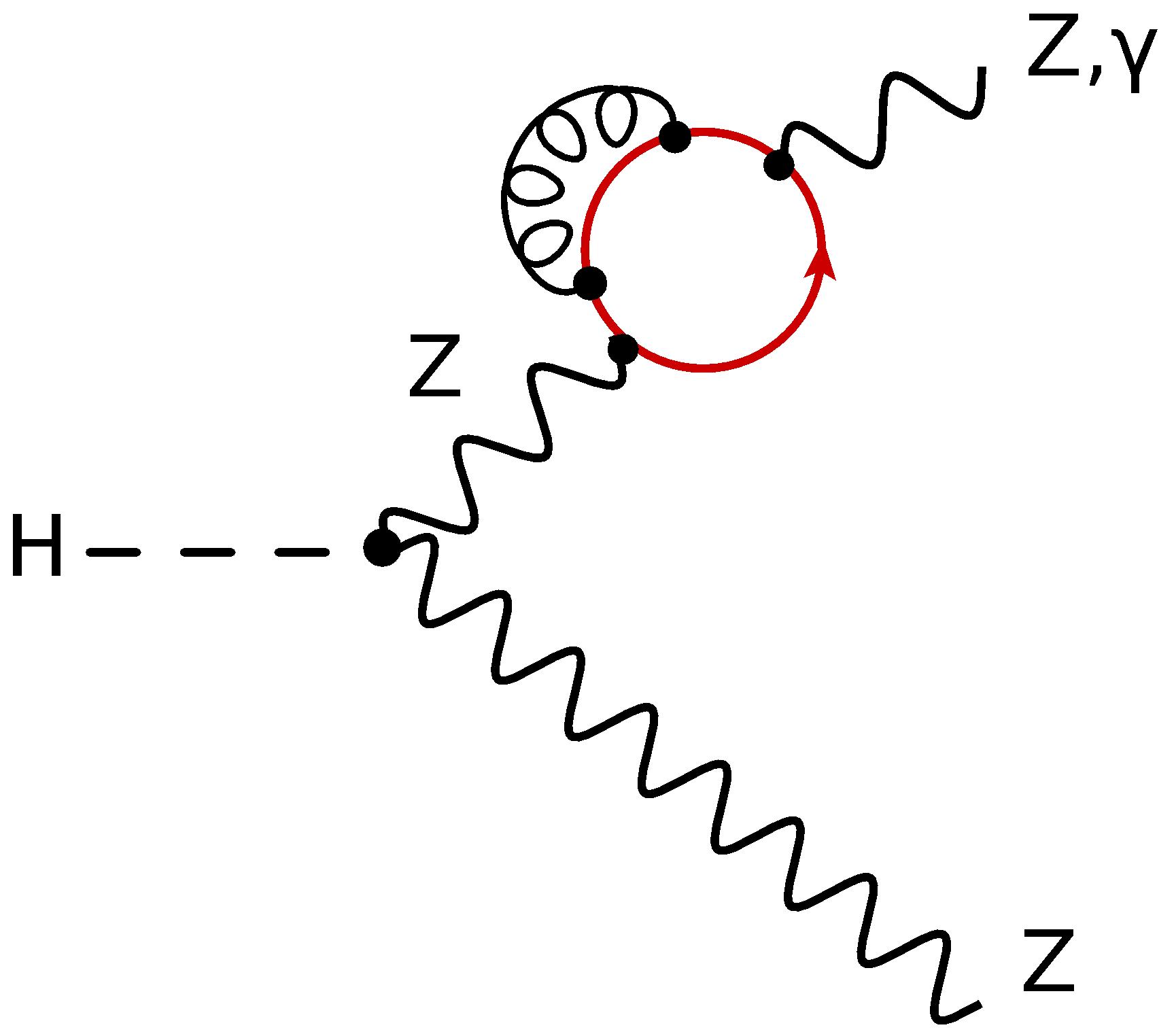}}
\hspace{1cm}
\subfloat[]{\includegraphics[width = 1.4in]{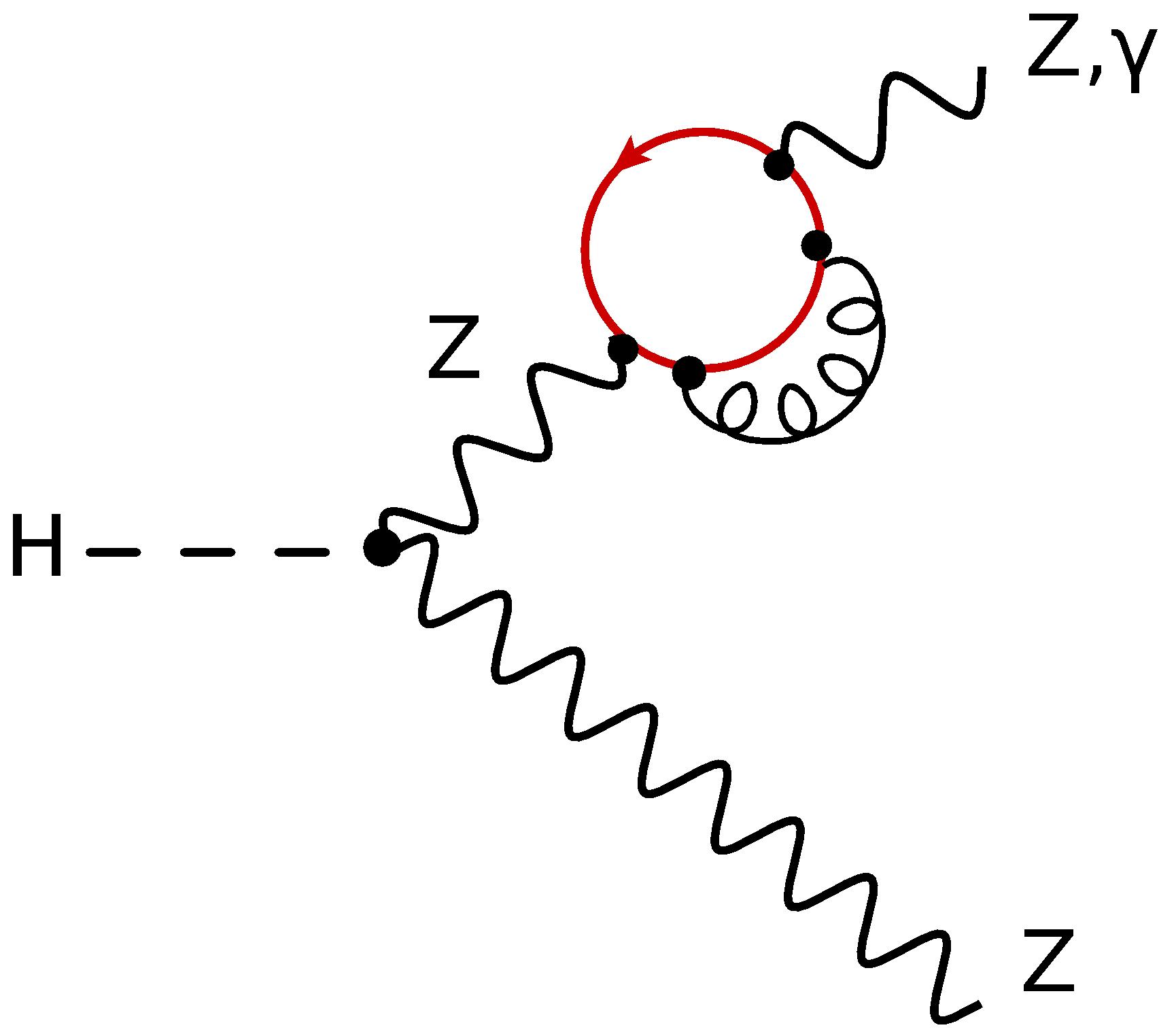}}\\
\caption{{{Representative diagrams contributing to $\delta M_{S.E.}^{\alpha \alpha_s}$ with a quark running in the loop. In these diagrams, five light quark flavors ($u, d, c, s$ and $b$) and the top quark contribute. The light quarks are taken to be massless.}}}
\label{fig:SE_bare}
\end{figure}

\subsection{$Z\ell \bar{ \ell}$  vertex correction} 
The amplitude at $\mathcal{O}(\alpha \alpha_s)$ also receives contribution from the diagrams shown in Fig.~\ref{fig:Zll_vertex} involving {a}} $Z\ell \bar{ \ell}$ vertex counterterm. 
The contribution to the vertex counterterm at $\mathcal{O}(\alpha \alpha_s)$ comes from the self-energies of the vector bosons. It depends on the 
wave function renormalization constants of the vector bosons, the charge renormalization
constant, and the renormalization constant for the weak mixing angle  \cite{Dittmaier:2015rxo}. The pure electroweak nature of the $Z\ell \bar{ \ell}$ vertex does not allow any kind of QCD corrections of $\mathcal{O}(\alpha \alpha_s)$; therefore, the counterterm for the $Z\ell \bar{ \ell}$ vertex is non-divergent in nature. One must be careful about the renormalization scheme in which the vertex correction is computed. 

\begin{figure}[ht!]
\centering
\subfloat[]{\includegraphics[width = 1.6in]{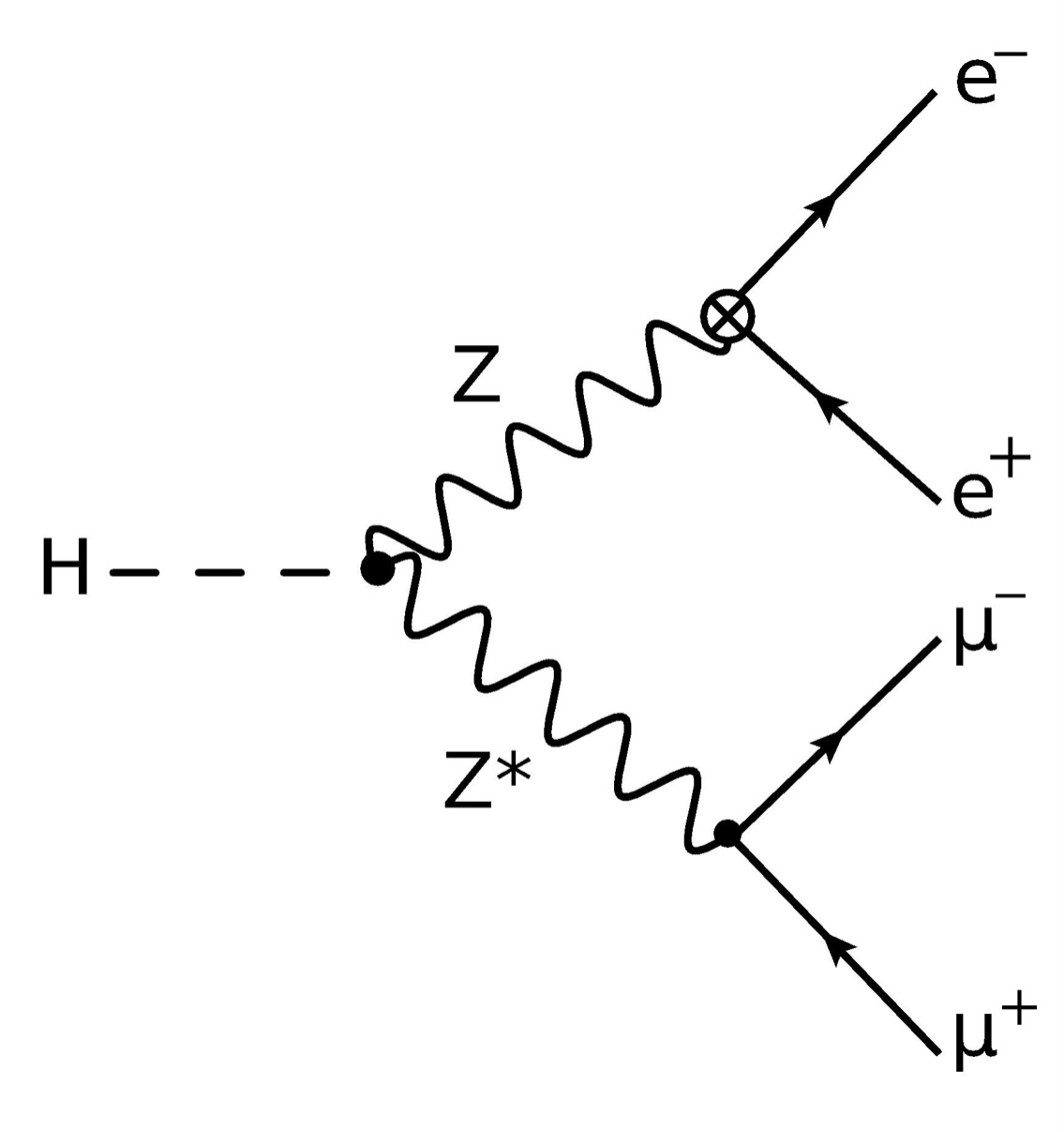}} 
\hspace{1cm}
\subfloat[]{\includegraphics[width = 1.6in]{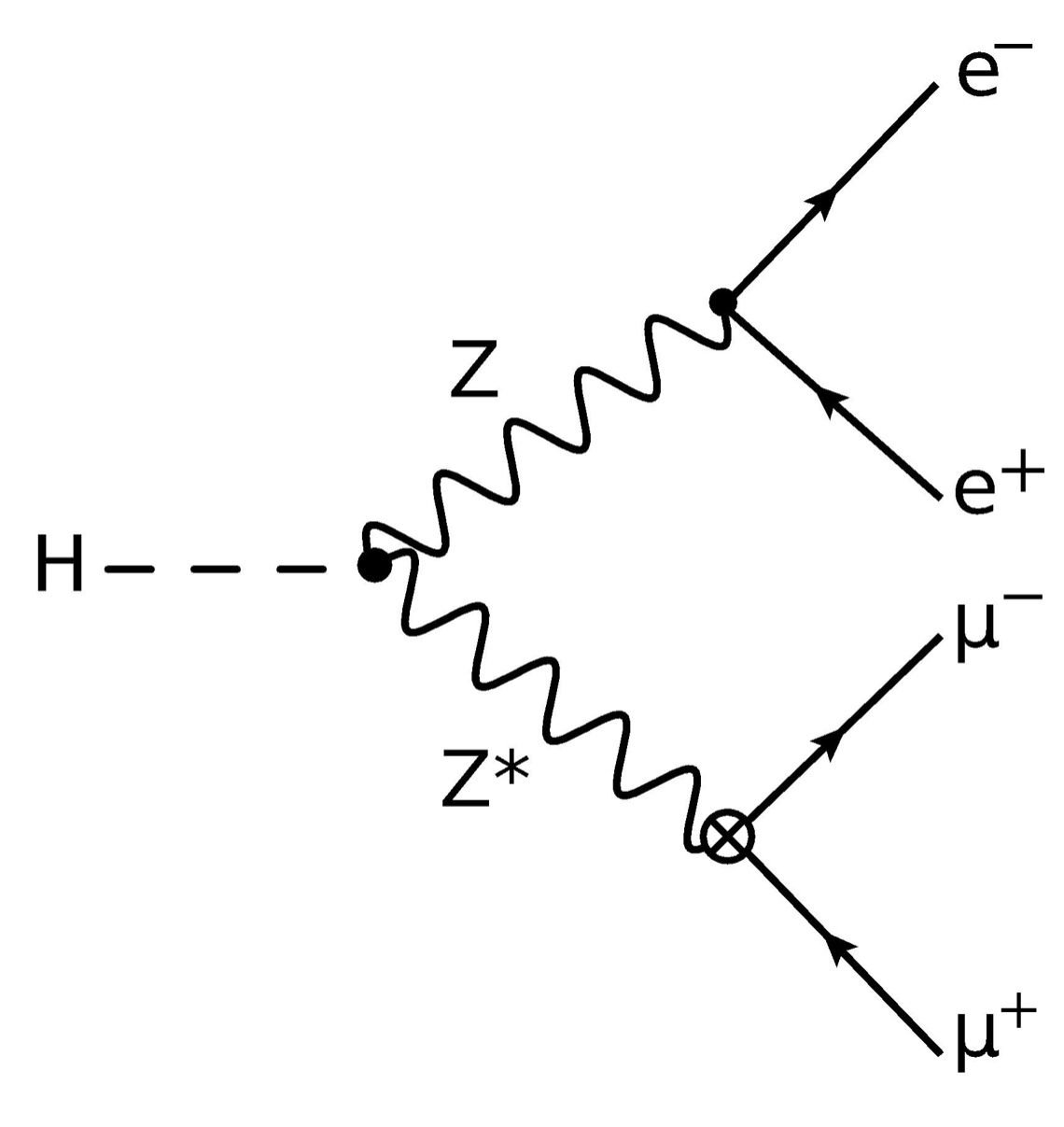}}
\caption{Tree level diagrams involving $\mathcal{O}(\alpha \alpha_s)$ $Z\ell \bar{ \ell}$ vertex counterterm denoted by a crossed circle.}
\label{fig:Zll_vertex}
\end{figure}

\section{Two-loop form-factors and divergences}
\label{sec:FF}
The form-factors for the $HV_1V_2$ vertex corrections {{given in Eq.~\ref{eqn:general amplitude}}} can be obtained by applying the projectors given in Appendix~\ref{app:A} on the bare two-loop amplitude of all the contributing diagrams shown in Fig.~\ref{fig:triangle}. 
In order to calculate the trace involving the $\gamma^5$ matrix, the prescription given in~\cite{Larin:1993tq} is used.
{Furry's theorem forbids the appearance of a single $\gamma^5$ in the trace over a closed fermionic loop due to charge invariance, which gives $C=0$ on adding the contributions from all the triangle diagrams together.} Furthermore, as the current conservation is associated with massless leptons in the final state, only the form-factors $A$ and $B$ contribute at the squared amplitude level. These form-factors can be written as linear combinations of scalar two-loop integrals of the type
\begin{align}
\label{eqn:integral-family}
 I_{\{\nu _i\}}\left( d,p_1^2, p_2^2, m_t^2, \mu^2 \right)
 & = 
 e^{2 \gamma_E \epsilon}
 \left(\mu^2\right)^{\nu-d}
 \int \frac{d^d k_1}{i \pi^{\frac{d}{2}}} \frac{d^d k_2}{i \pi^{\frac{d}{2}}}
 \prod\limits_{i=1}^7 \frac{1}{ P_i^{\nu_i} }.
\end{align}
Here, $k_1$ and $k_2$ are the loop momenta, $d$ denotes the space-time dimensions, $\gamma_E$ is the Euler-Mascheroni constant, $\mu$ represents an arbitrary scale introduced to maintain the canonical dimension of the integral, $\{\nu _i\} $ = $\{\nu_1 \nu_2 \nu_3 \nu_4 \nu_5 \nu_6 \nu_7\}$ are powers of the inverse propagators $P_i$, and $\nu = \sum_{i=1}^7 \nu_i$.
The inverse propagators $P_i$ are given by
\begin{align}
\label{eqn:propagators}
 P_1 & = k_1^2 - m_t^2,
 &
 P_2 & = k_2^2 - m_t^2,
 &
 P_3 & = \left(k_1-k_2\right)^2,
 \nonumber \\
 P_4 & = \left(k_1-p_1\right)^2 - m_t^2,
 &
 P_5 & = \left(k_2-p_1\right)^2 - m_t^2,
 &
 P_6 & = \left(k_1-p_1-p_2\right)^2 -m_t^2,
 \nonumber \\
 P_7 & = \left(k_2-p_1-p_2\right)^2 - m_t^2.
\end{align}

The set of two-loop integrals in these form-factors is reduced to a minimal set of integrals called master integrals (MIs) using the integration-by-parts (IBP)~\cite{Tkachov:1981wb,Chetyrkin:1981qh} and Lorentz-invariance (LI) \cite{Gehrmann:1999as} identities with the programs \texttt{LiteRed}~\cite{Lee:2012cn,Lee_2014} combined with \texttt{Mint}~\cite{Lee:2013hzt} and \texttt{FIRE}~\cite{A.V.Smirnov_2008,SMIRNOV20132820,SMIRNOV2015182}. We find a total of 41 master integrals after the IBP reduction. The choice of master integrals is not unique. The basis set $\vec{I}$ of 41 master integrals we obtain is listed below.
\begin{equation*}
\begin{aligned}
& \{I_{0000011},\; I_{0000111},\; I_{0001111},\; I_{0010110},\; I_{0020110},\; I_{0100011},\; I_{0100110},\; I_{0100111},\; I_{0101011},\; I_{0101110},\\
& I_{0101111},\; I_{0110010},\; I_{0110110}, \;I_{0111000},\; I_{0111001},\; I_{0111010},\; I_{0111011}, \;I_{0111110},\; I_{0112011},\; I_{0112110},\\
& I_{0120110},\; I_{0121001},\; I_{0121010},\; I_{0121011},\; I_{0121110},\; I_{0210010}, \;I_{0210110},\; I_{0211000},\; I_{0211001},\; I_{0211010},  \\
& I_{0211011},\; I_{0211110},\; I_{1100011},\; I_{1100110},\; I_{1100111},\;
I_{1101100},\; I_{1101101},\; I_{1110110},\; I_{1120110},\; I_{1210110}, \\
& I_{2110110}\}.
\end{aligned}
\end{equation*}

The involvement of two-loop integrals makes the evaluation of these form-factors highly non-trivial. The master integrals can be evaluated numerically with the help of publicly available codes such as pySecDec~\cite{BOROWKA2018313} {{and}} AMFlow~\cite{Liu:2022chg} etc. The analytical results for these 41 master integrals in the canonical basis are now also available in terms of iterated integrals~\cite{Chaubey:2022hlr}. {In our work, for an efficient evaluation, keeping the required accuracy of the MIs in mind, we use an in-house code for the numerical evaluation of all the two-loop master integrals involved, using the sector-decomposition method given in~\cite{Binoth:2000ps,Binoth:2003ak,Li:2015foa}. The results obtained for the MIs using the numerical integration are in good agreement with the analytical results.}

Due to the presence of loop integrals and massless particles in the loop, the two-loop form-factors develop both the Ultraviolet (UV) and Infrared (IR) divergences. 
We regularize these divergences in dimensional regularization by taking $d = 4-2\epsilon$. After regularization, the divergences are encoded in the two-loop master integrals as poles in $\epsilon$, $\dfrac{1}{\epsilon^4}$ being the highest order of pole that can 
appear. The $\dfrac{1}{\epsilon^3}$ and $\dfrac{1}{\epsilon^4}$ poles are exclusively 
due to the IR singularities, while $\dfrac{1}{\epsilon}$ and $\dfrac{1}{\epsilon^2}$ poles
can be due to both IR and UV singularities. According to the KLN (Kinoshita-Lee-Nauenberg) theorem~\cite{Kinoshita:1962ur,PhysRev.133.B1549,PhysRevD.17.2773,PhysRevD.17.2789}, the IR singularities eventually get cancelled against the real emission Feynman diagrams to give IR safe observables. 

\begin{figure}[!h]
\begin{center}
\includegraphics[width=5cm,height =2.8cm]{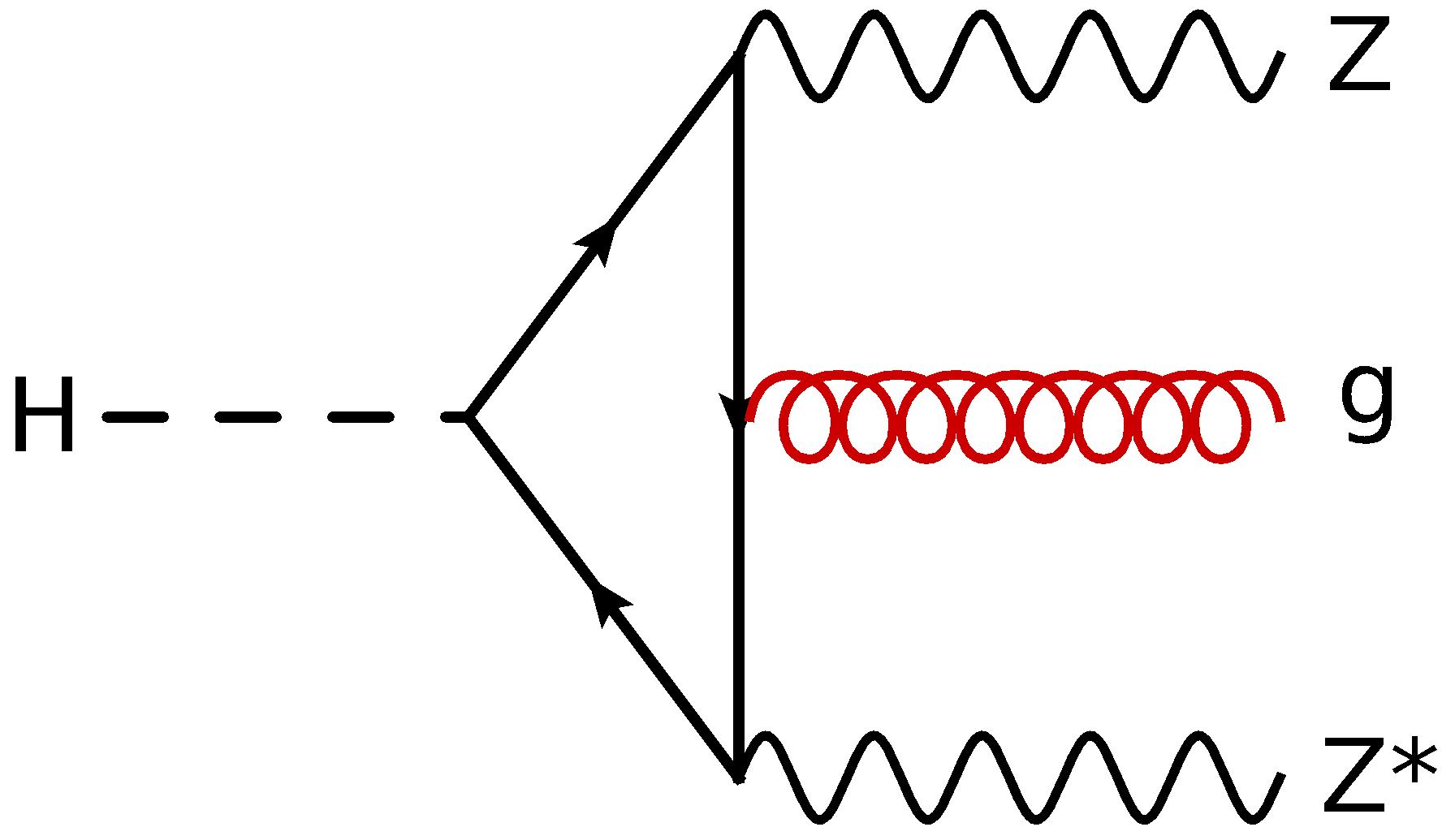}
\end{center}
\caption{Representative diagram for real corrections to the amplitude for $H \to ZZ^*$.}
\label{fig:IR}
\end{figure}

{In our case, the possible real emission Feynman diagrams involve the emission of a gluon from the closed quark loop, as shown in Fig.~\ref{fig:IR}.}
{Due to a closed fermionic loop, the amplitudes of these diagrams are proportional to the trace over $T^a$, the generator of the $SU(3)$ gauge group. Since ${\rm tr}(T^a)=0$, diagrams for the real corrections give zero. Moreover, these real emission diagrams can contribute only at the amplitude-squared level, which is an $\mathcal{O}(\alpha^2 \alpha_s)$ effect with respect to the LO. 
Therefore, the real emission diagrams do not contribute at $\mathcal{O}(\alpha \alpha_s)$. The absence of real corrections thus demands the cancellation of virtual IR divergences among the contributing diagrams, and the two-loop amplitude is expected to be free from any IR divergences in our case. Since the form-factors $A$ and $B$ for the $HV_1V_2$ vertex are independent, at the two-loop, they should be separately free from $\dfrac{1}{\epsilon^3}$ and $\dfrac{1}{\epsilon^4}$ poles. This fact will provide one of the important checks on our calculation.} Furthermore, since the form-factor $B$ is zero at the tree-level, and the first non-zero contribution to it arises at the one-loop, we expect that the two-loop form-factor $B$ does not have $\dfrac{1}{\epsilon^2}$ UV pole dependence. This can serve as another consistency check on our calculation. 

\section{UV renormalization and the complex-mass scheme}
\label{sec:renorm}
{In order to remove the UV divergences from the matrix elements contributing at $\mathcal{O}(\alpha \alpha_s)$, the standard on-shell renormalization scheme is used to renormalize all the fields and masses involved. The renormalization process involves evaluating the required counterterm (CT) diagrams and then adding the CT amplitude back to the UV divergent amplitude to obtain finite results after the cancellation of all the divergences}. In total, there are 48 one-loop triangle, 96 one-loop self-energy, and 8 tree-level CT diagrams contributing to the amplitude at $\mathcal{O}(\alpha \alpha_s)$. The representative CT diagrams are shown in Figs.~\ref{fig:triangleCT},~\ref{fig:SE_CT} and~\ref{fig:Tree_CT}. 

As shown in Fig.~\ref{fig:triangleCT}, the triangle counterterm diagrams mainly involve the $Vt\bar{t}$, $Ht\bar{t}$ vertex counterterms and counterterms for the top-quark mass and wave function. However, the renormalization of the quark wave function is related to the vertex renormalization. As a result, the vertex counterterms involving quarks completely cancel out the contributions from the quark wave function counterterms. {{It is not surprising, as in our calculation, the total amplitude does not depend on the quark wave function. Thus, the complete cancellation of the quark wave function counterterm provides another important check on our calculation}. Therefore, at the one-loop level, we only need to evaluate diagrams with the top-quark mass counterterm insertions.
On the other hand, for the evaluation of the self-energy and tree-level counterterm diagrams, one needs $\mathcal{O}(\alpha \alpha_s)$ expressions for {the renormalization constants $\delta{Z_e}$, $\delta{Z_{ZZ}}$, $\delta{Z_H}$, $\delta{M_Z^2}$, $\delta{M_W^2}$, $\delta{Z_{\gamma Z}}$ and $\delta{Z_{Z\gamma}}$, which we have deduced from the self-energies of {the Higgs and gauge bosons given in~\cite{Djouadi:1993ss,PhysRevD.51.218}.

\begin{figure}[ht!]
\centering
\subfloat[]{\includegraphics[height=0.9in,width = 1.4in]{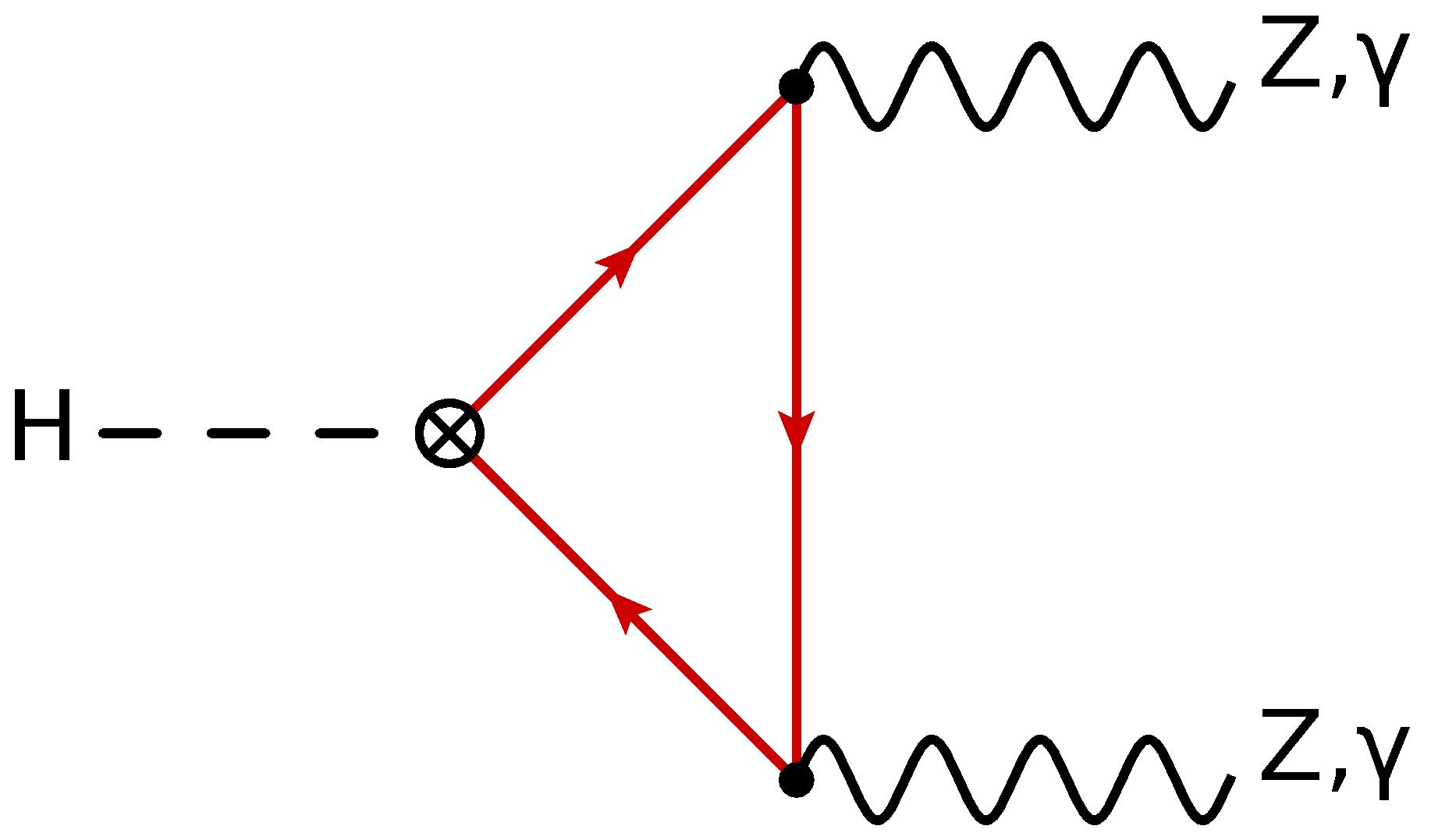}} 
\hspace{1cm}
\subfloat[]{\includegraphics[height=0.9in,width = 1.4in]{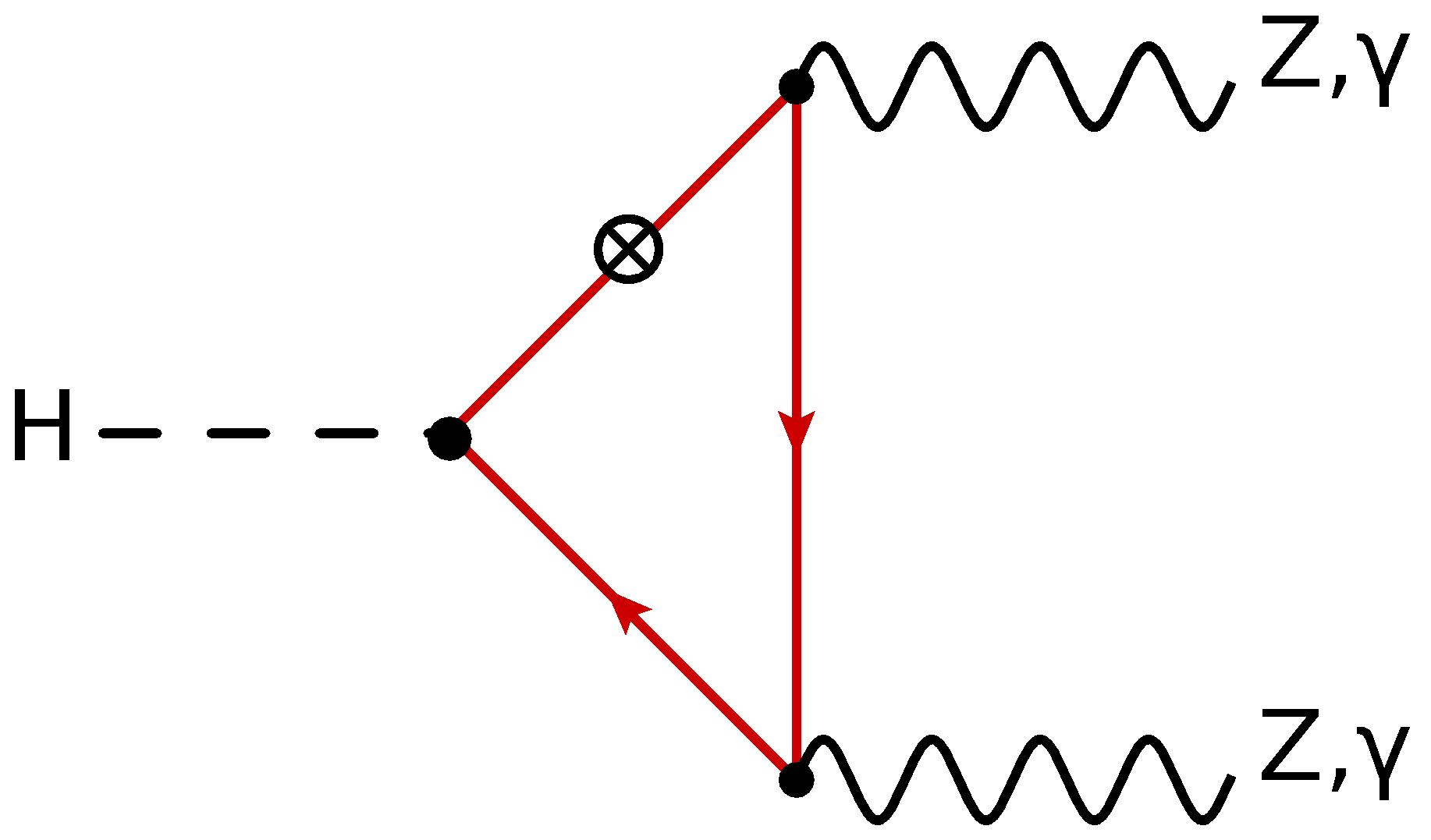}}
\hspace{1cm}
\subfloat[]{\includegraphics[height=0.9in,width = 1.4in]{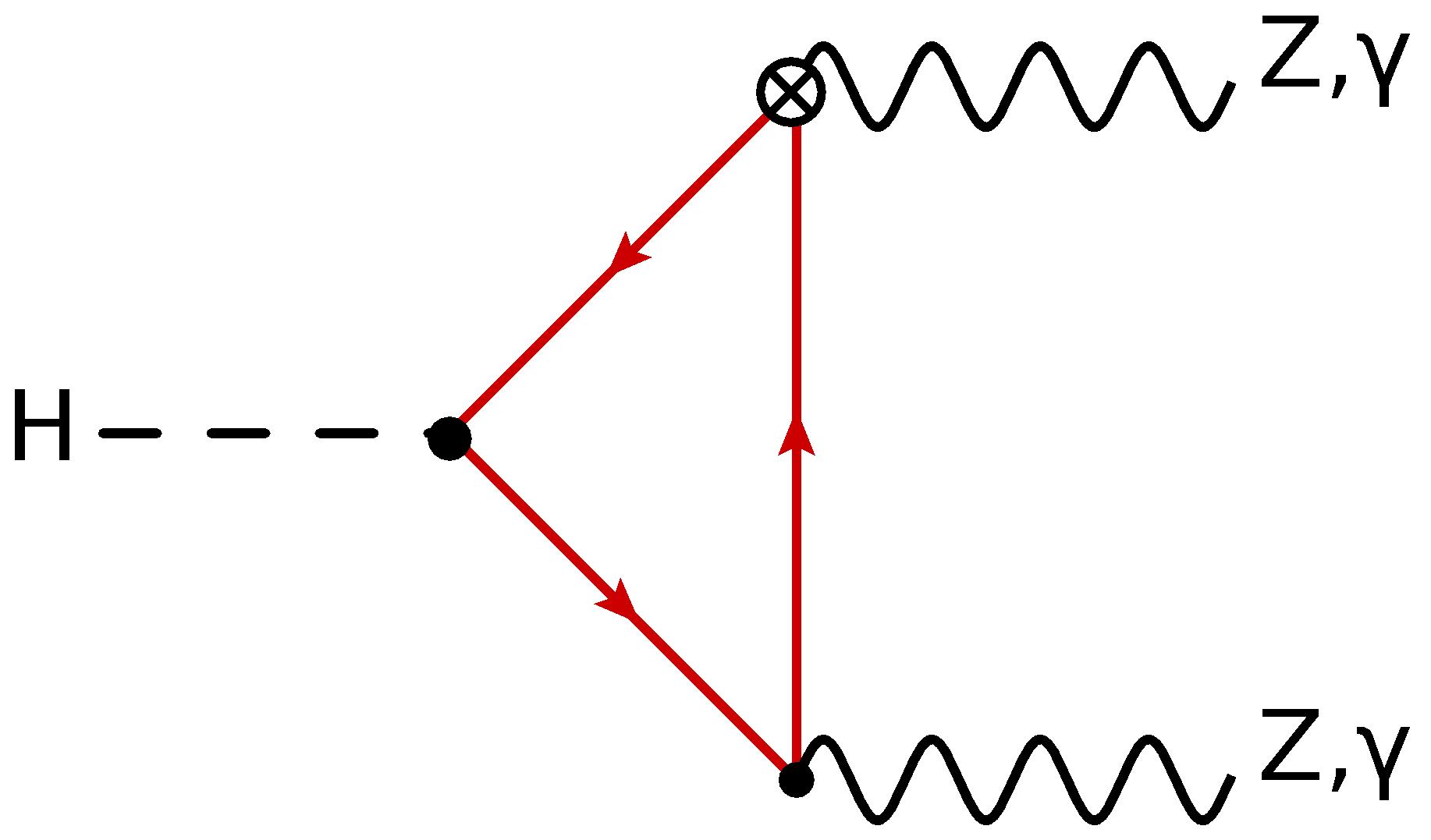}} 
\caption{Representative one-loop triangle CT diagrams. The counterterm vertex proportional to $\alpha_s$ is denoted by a crossed circle.}
\label{fig:triangleCT}
\end{figure}
\begin{figure}[ht!]
\centering
\subfloat[]{\includegraphics[width = 1.2in]{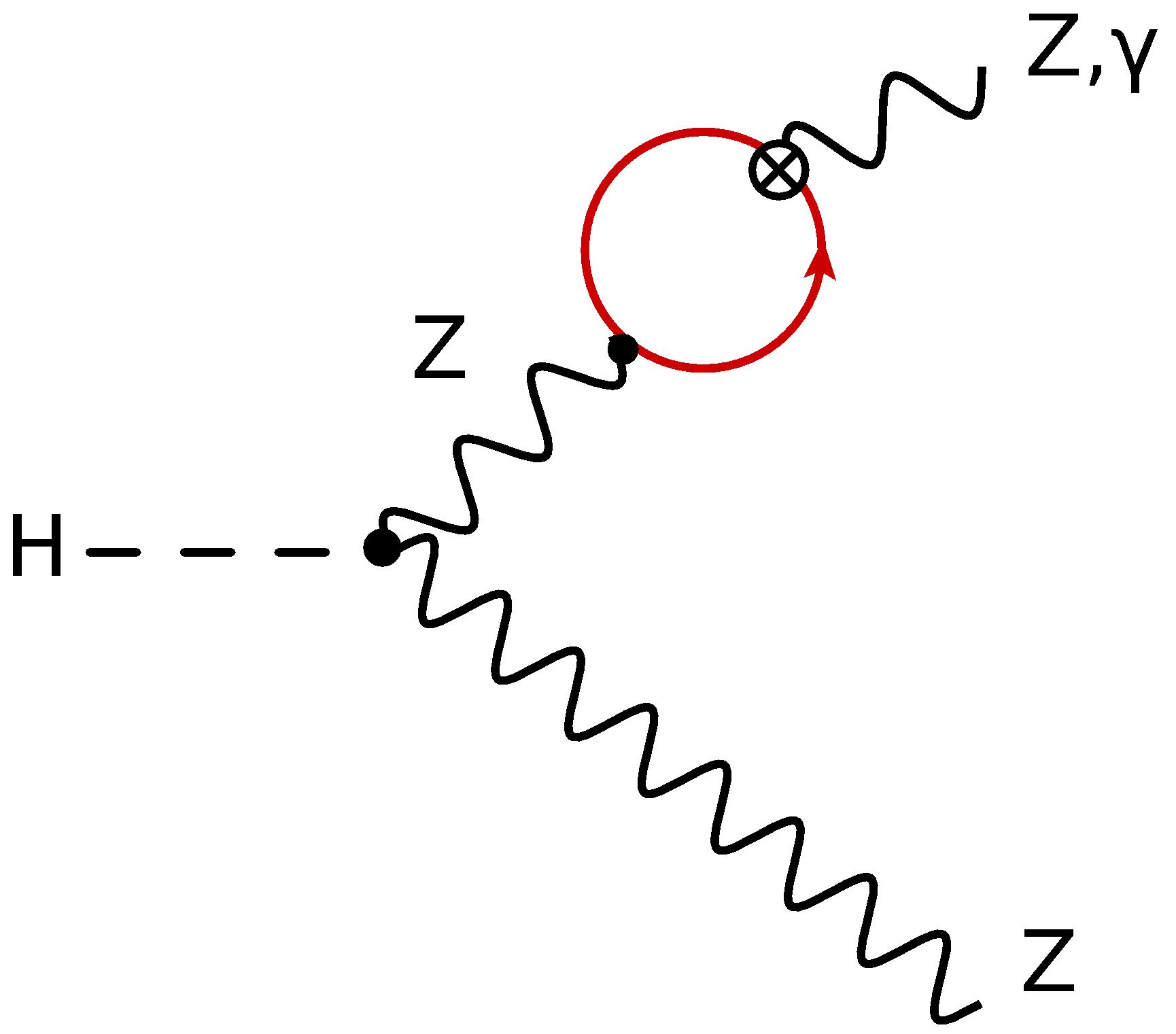}} 
\hspace{0.5cm}
\subfloat[]{\includegraphics[width = 1.2in]{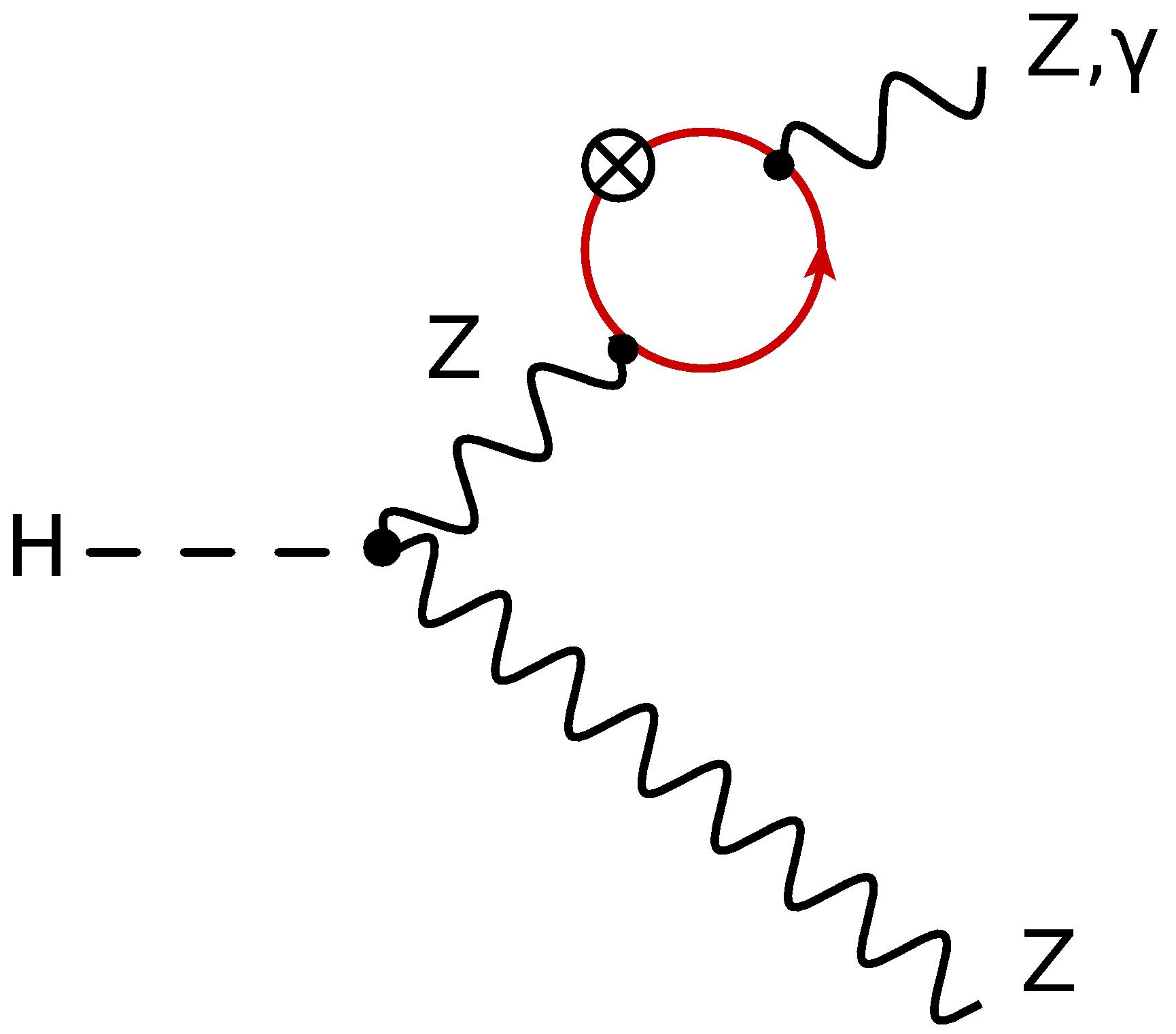}}
\hspace{0.5cm}
\subfloat[]{\includegraphics[width = 1.2in]{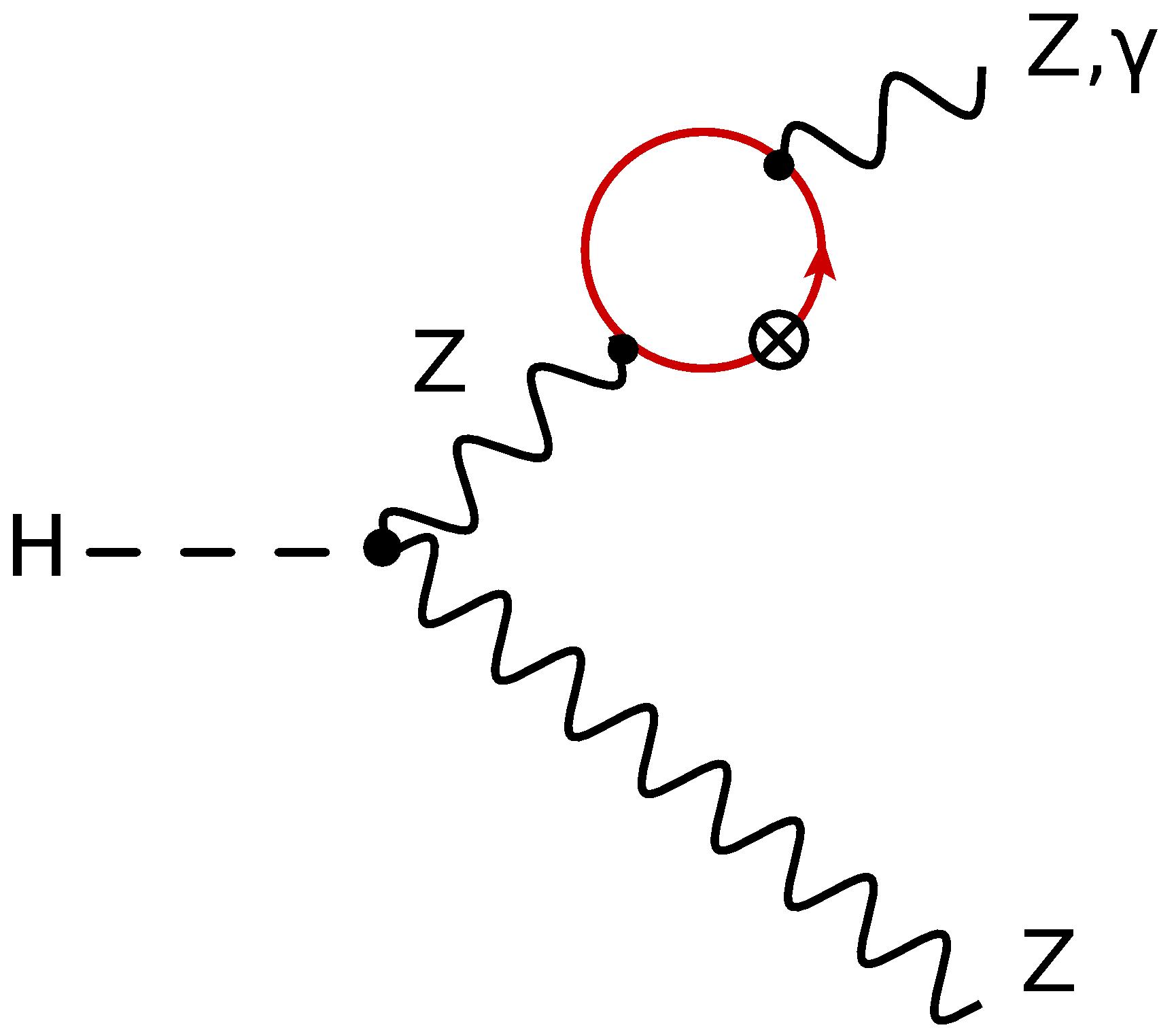}} 
\hspace{0.5cm}
\subfloat[]{\includegraphics[width = 1.2in]{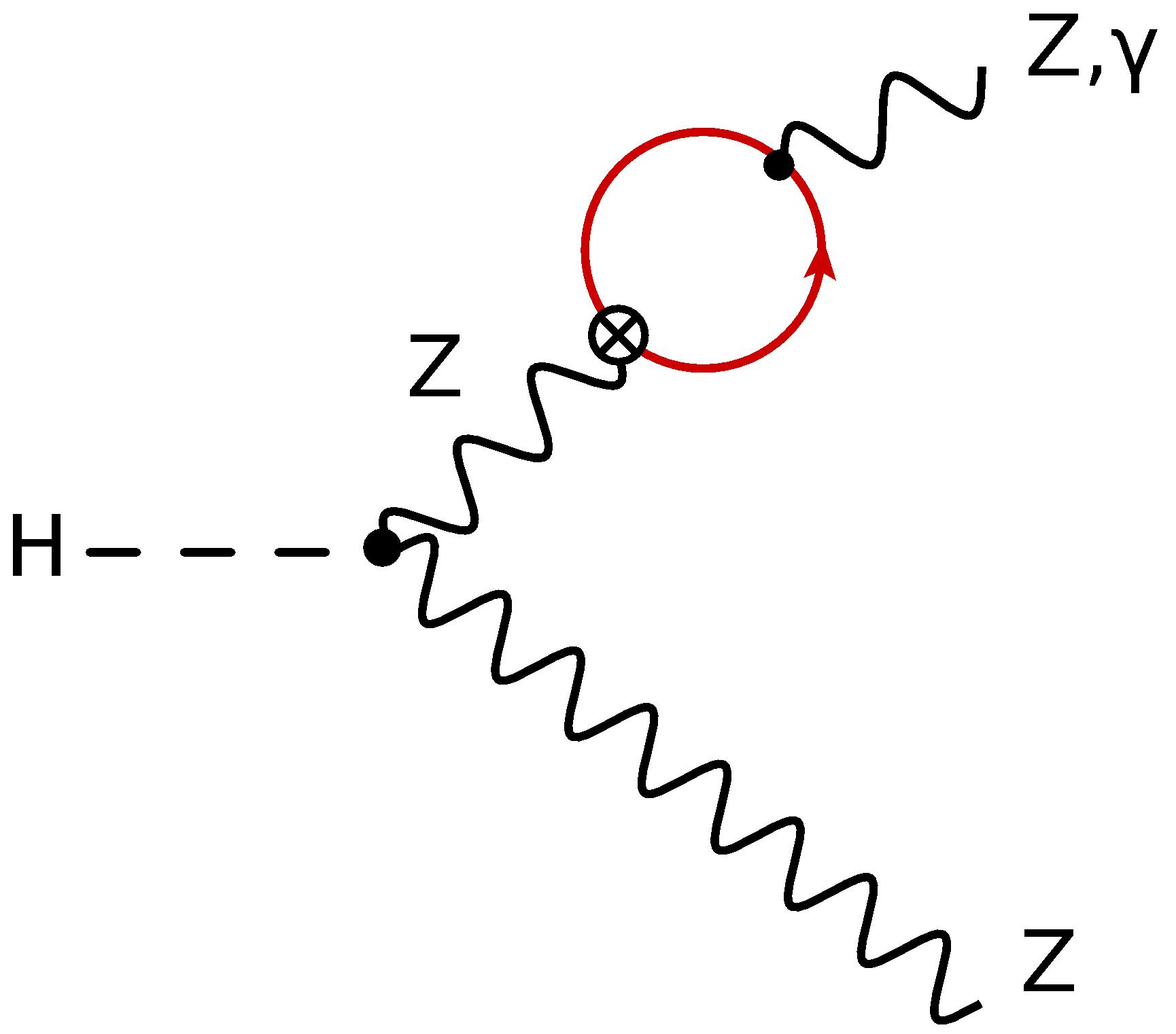}} 
\caption{Representative one-loop self-energy CT diagrams. The counterterm vertex proportional to $\alpha_s$ is denoted by a crossed circle. The diagrams with counterterm insertions on the lower leg are not shown.}
\label{fig:SE_CT}
\end{figure}
\begin{figure}[ht!]
\centering
\subfloat[]{\includegraphics[width = 1.3in]{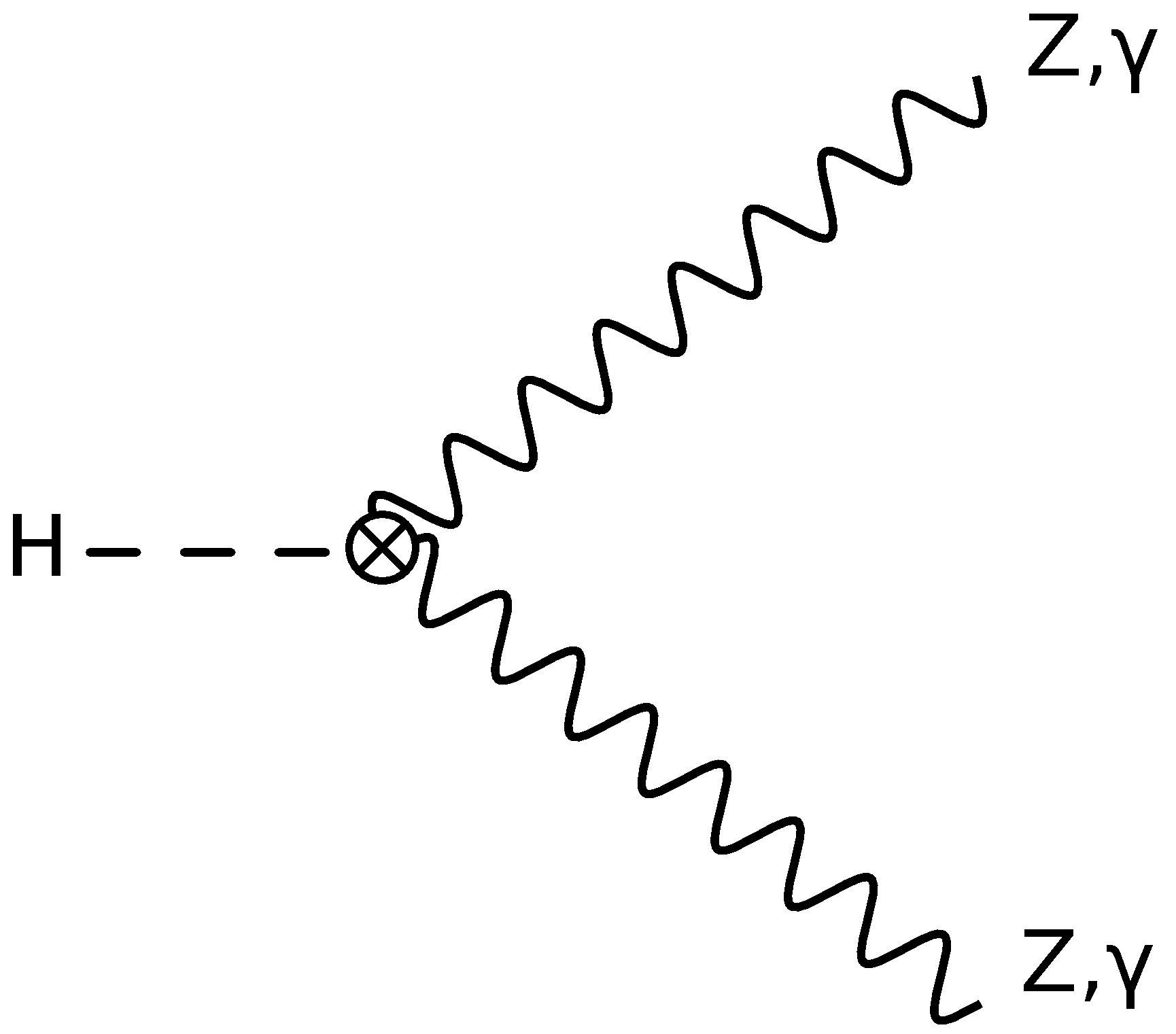}} 
\hspace{1cm}
\subfloat[]{\includegraphics[width = 1.3in]{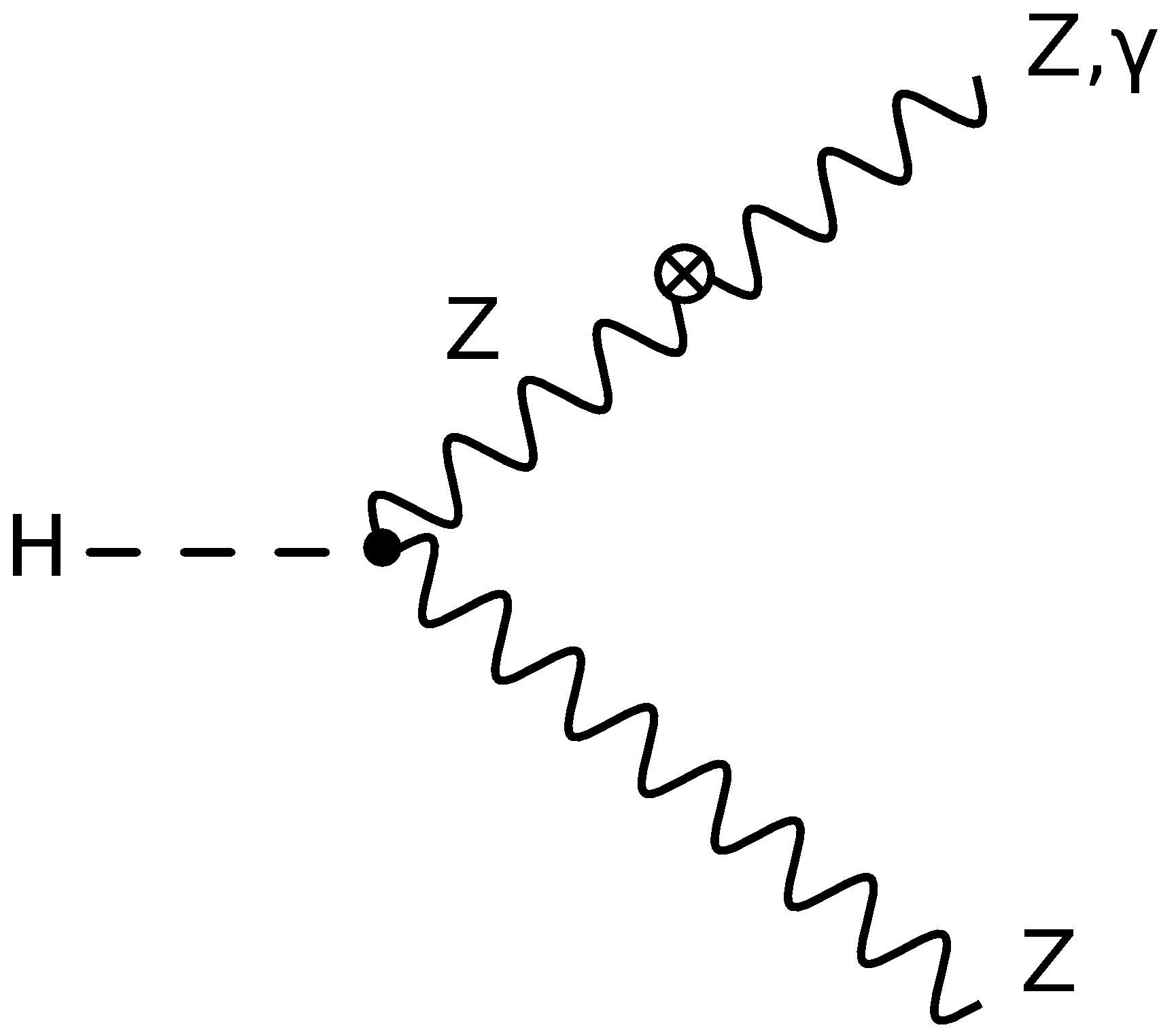}}
\hspace{1cm}
\subfloat[]{\includegraphics[width = 1.3in]{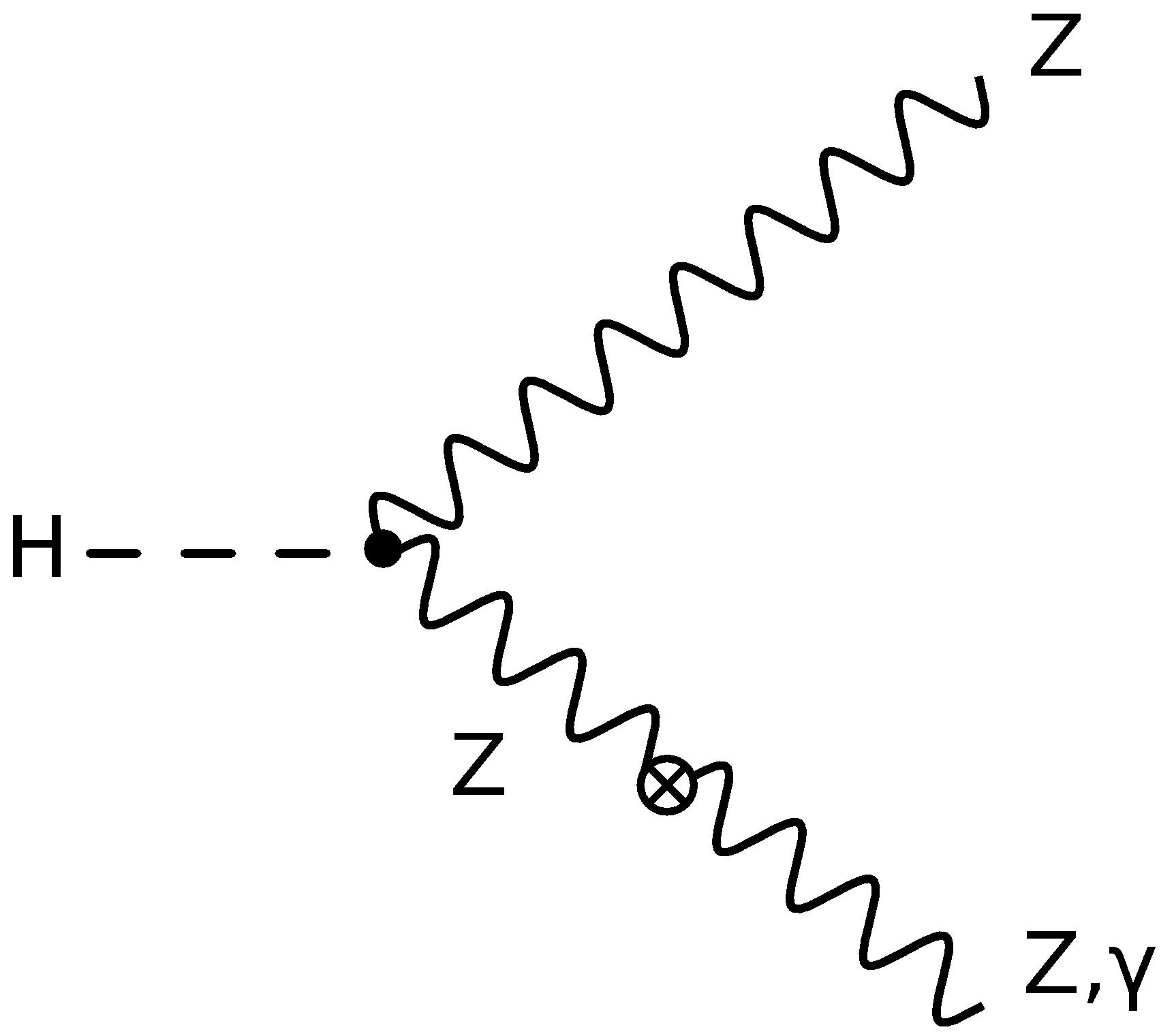}}
\caption{Tree level CT diagrams. The counterterm vertex proportional to $\alpha \alpha_s$ is denoted by a crossed circle.}
\label{fig:Tree_CT}
\end{figure}

{The diagrams for the process under consideration involve unstable $Z$-bosons in the propagators. Therefore, one needs to introduce the finite $Z$-width in the propagators to ensure the stability of the perturbative calculation at the $Z$-pole.} However, this incorporation of the finite $Z$-width can lead to several problems, such as the violation of gauge invariance due to the mixing of different perturbative orders~\cite{Grunewald:2000ju}. These problems are tackled via the adoption of the ``complex-mass scheme" (CMS)~\cite{Denner:1999gp,Denner:2005es,Denner:2005fg,DENNER200622}. {{In this scheme, we analytically continue the masses of the weak gauge bosons to the complex plane as $M_V^2 \to \; M_V^2 - i M_V {\rm{\Gamma_V}}$, where, $V=W,Z$ and ${\rm{\Gamma_V}}$ is the corresponding {constant} decay width. This scheme also makes the weak mixing angle complex as $\cos^2\theta_W= (M_W^2 - i M_W{\rm{\Gamma_W}})/ (M_Z^2 - i M_Z {\rm{\Gamma_Z}}$).}} Therefore, the CMS scheme is just a generalization of the on-shell scheme with complex renormalization constants. There is no need to introduce a finite width for the top quark running in the loop, as the Higgs mass is taken to be 125 GeV, which is below the $t\bar{t}$ threshold. 

After combining the CT amplitude with the UV divergent amplitude in the complex mass scheme, we get the finite results for the amplitudes of the $HV_1 V_2$ vertex and the self-energy corrections.

\section{Numerical implementation and checks}
\label{sec:imple}
We combine the UV finite amplitudes for the $HV_1 V_2$ vertex and self-energy corrections with the fermionic currents to get the two-loop amplitude for $H \to e^+ e^-\mu^+\mu^-$. Adding the finite $Z\ell \bar \ell$ contribution to it, 
we get the matrix element $M_2^{\alpha \alpha_s}$ given in Eq.~\ref{eqn:Total_Amp}.
In the perturbative expansion of the amplitude-squared up to two-loop, this matrix element interferes with the LO amplitude as
\begin{equation}
|M|^2 = |M_0|^2 + 2\;Re(M_0^*\; M_1^{\alpha}) + 2\;Re(M_0^*\; M_2^{\alpha \alpha_s}).
\end{equation} 
The interference term is organized in terms of the two-loop form-factors using the symbolic manipulation program \texttt{FORM}, and a {\tt FORTRAN} output 
is obtained for the numerical evaluation. To obtain the partial decay width, we need to perform the phase-space integration over the final state leptons. This is done using the publicly available {\tt{Hto4l}}~\cite{Boselli:2015aha} code, which is a Monte Carlo code that generates events for the process $ H \to 4\ell\;(\ell=e,\mu)$. 
To achieve {a} good accuracy on the observables of interest, we need to sample a huge number of phase-space points. However, 
calculating the form-factors for every phase point is very time-consuming. To manage this issue, {we have prepared a two-dimensional grid for the form-factors $A$ and $B$ using pre-specified phase-space points ($p_1^2$ and $p_2^2$ values to be more precise) by numerically evaluating 41 MIs for input parameters given in section~\ref{sec:results}. The grid is prepared with an accuracy of $\mathcal{O}(10^{-3})$. 
The grid is then used to estimate the form-factors at random phase-space points with the 
help of a linear interpolation code developed in-house. It is important to note that any change in the input parameter set would require a new grid.}
We interface the squared matrix elements, the grid of the form-factors, and the interpolation code with the {\tt{Hto4l}} code to perform the phase-space integration. This allows us to obtain the partial decay width and kinematical distributions for the final state leptons at ${\cal O}(\alpha \alpha_s)$. In order to prove the reliability of our implementation, we have performed the following checks:
\begin{enumerate}
    \item To good numerical accuracy, we find that the $\dfrac{1}{\epsilon^4}$ and $\dfrac{1}{\epsilon^3}$ poles cancel in both the form-factors $A$ and $B$. In the form-factor $B$, the 
         $\dfrac{1}{\epsilon^2}$ pole also vanishes. The UV poles in the form-factors $A$ 
         and $B$ cancel after adding the CTs, and the result does not depend on the choice of the scale $\mu$ in the dimensional regularization. These checks have been performed for several phase-space points. 

         \item {{By taking the gluon propagator in the $R_\xi$ gauge, we find that the two-loop form-factors, and consequently the two-loop amplitude, {{are}} gauge-parameter independent. Additionally, the Ward identity for the $HV_1 V_2$ vertex demands $A + p_1.p_2\;B + p_1^2\; D = 0$, and we have verified this relation for the two-loop amplitude at $\mathcal{O}(\alpha \alpha_s)$.}}
         
    \item {As mentioned earlier, the two-loop diagrams for $H \to e^+e^-\mu^+\mu^-$ are closely related to the ones appearing in the production process $e^+ e^- \to ZH$. In Ref.~\cite{Gong:2016jys}, analytical expressions for the contributing form-factors are given up to order $m_t^{0}$ after series expanding them in $\frac{1}{m_t}$. In order to check the accuracy of the grid prepared for the form-factors, we produced the grid for $e^+ e^- \to Z H $, taking a very large value of the top-quark mass ($m_t$). Further, we matched the numerical values of the form-factors from the grid with those given in~\cite{Gong:2016jys}. We found an excellent agreement between the two for different values of the center-of-mass energies.}
    
     \item The correctness of our numerical implementation is checked via reproducing the results for the mixed QCD-electroweak corrections for the $e^+ e^- \to ZH$ process given in Ref.~\cite{Sun:2016bel} in the $G_\mu$ and $\alpha(0)$ schemes. We performed this check by implementing our calculation in \texttt{MadGraph} \cite{Alwall:2014hca}, and we found that the calculated corrections matched the available results in both the schemes with a relative error of less than $1\%$.
\end{enumerate}

\section{Numerical results}
\label{sec:results}

In this section, we will present the numerical results for our calculation obtained by its implementation in the {\tt{Hto4l}} code. We work in the $G_{\mu}$ scheme and use the following set of input parameters,
\begin{equation} 
\begin{aligned}
G_\mu  &= 1.1663787\times 10^{-5}\; {\rm GeV}^{-2}, & M_Z^{\rm OS}  &= 91.1876 \;{\rm GeV},& M_W^{ \rm OS}  &= 80.379 \;{\rm GeV}, \\
\Gamma_Z^{\rm OS}&= 2.4952 \;{\rm GeV},& \Gamma_W^{ \rm OS}&= 2.141 \;{\rm GeV},&M_H  &= 125 \;{\rm GeV},\\ m_t &= 173\;{\rm GeV},& \alpha_s(M_Z)& = 0.1185.
\end{aligned}
\end{equation}
In the $G_\mu$ scheme, $\alpha$ is given by 
\begin{equation}
\alpha = \frac{\sqrt{2}G_\mu M_W^2}{\pi}\bigg(1-\frac{M_W^2}{M_Z^2}\bigg),
\end{equation}
which is scale-independent in the case of on-shell renormalization. {From the above list of input parameters, the transformation of the on-shell values of the masses and widths ($M_V^{\rm OS}, \Gamma_V^{\rm OS}$) of the $W$ and $Z$ bosons to the corresponding values in the CMS, denoted by $M_V$ and $\Gamma_V$, is provided in Appendix~\ref{app:B}. It is worth mentioning that the difference between using $(M_V^{\rm OS}, \Gamma_V^{\rm OS})$ or $(M_V, \Gamma_V)$ is hardly visible in the numerical results presented in this section.} The results for the mixed QCD-electroweak corrections are obtained using both fixed and running $\alpha_s(Q)$. We take $Q=M_Z$ to obtain 
results with a fixed value of the QCD coupling. 
Since the mixed QCD-electroweak corrections are leading order in $\alpha_s$, we employ the one-loop result for the running of $\alpha_s(Q)$.
For the choice of the running scale, we have used $p_T$ of the lepton and the invariant mass of the $\ell^+\ell^-$ pair ($M_{\ell^+ \ell^-}$).

We find that the mixed QCD-electroweak correction
to the partial decay width is around 0.27$\%$ of the LO contribution for $\alpha_s$ at $Q=M_Z$\footnote{For any fixed scale choice $Q$ other than $M_Z$, the \% correction can be estimated by multiplying 0.27 with a factor of $\alpha_s(Q)/\alpha_s(M_Z)$.}. With the running coupling, it becomes 0.30\% for $Q=M_{\ell^+ \ell^-}$ and 0.35\% for $Q=p_T(\ell^-)$. 
To draw a comparison, we note that the inclusive two-loop QCD corrections in $H \to Z\gamma$ decay have been found around 0.22\% of the LO~\cite{SPIRA1992350,Gehrmann:2015dua,Bonciani:2015eua}. The two-loop QCD corrections in the $H \to \gamma \gamma$ decay lie in the range of 1-2\% for the intermediate Higgs mass below the $t\bar t$ threshold~\cite{DJOUADI1991187}. With respect to the NLO EW correction, the mixed QCD-electroweak correction amounts to 18\% for the fixed and 21\% (24\%) for the running QCD coupling with $Q=M_{\ell^+ \ell^-}$ ($Q=p_T(\ell^-)$). 

It is well known that the higher-order corrections are sensitive to the kinematics of the events. In Fig.~\ref{fig:dist1}, we investigate the impact of the two-loop corrections with respect to the LO predictions and NLO EW corrections on the invariant mass distribution of the final state lepton pair. We define $\delta_i = {\rm \Gamma_{\textit{two-loop}} / \Gamma}_{i}$ ($i=$ LO, NLO) to indicate the relative correction with respect to the LO contribution and NLO EW correction.
\begin{figure}[ht!]
\centering
\subfloat[]{\includegraphics[width=0.6\linewidth]{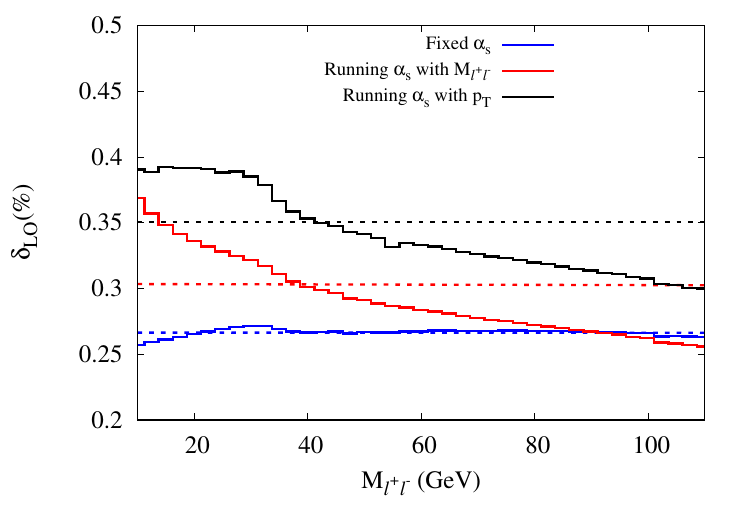}}
\newline
\centering
\subfloat[]{\includegraphics[width=0.5\linewidth]{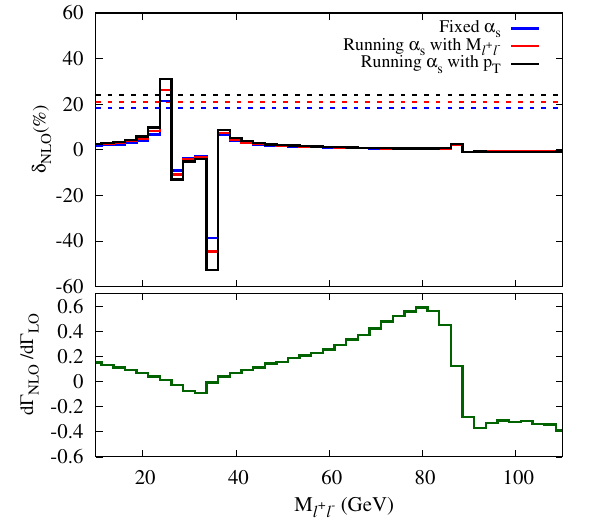}} 
\subfloat[]{\includegraphics[width=0.5\linewidth]{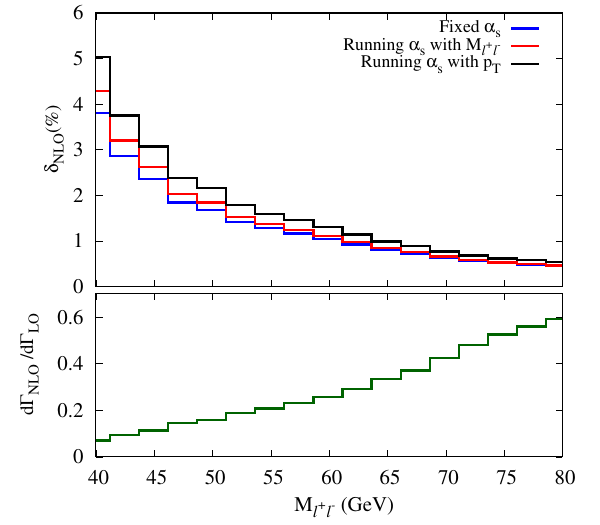}}
\caption{Effect of the mixed QCD-electroweak corrections on the invariant mass distribution of the final state lepton pair $\ell^+\ell^-$. 
In (a), $\delta_{\rm LO}$ is the ratio of the mixed QCD-electroweak correction and the LO contribution. In the upper panels of (b) and (c), $\delta_{\rm NLO}$ is the ratio of the mixed QCD-electroweak correction and the NLO EW correction. The quantity $\rm {d\Gamma_{\rm NLO}/d\Gamma_{\rm LO}}$ in the lower panels of (b) and (c) gives the ratio of the NLO EW correction and the LO contribution.
For clarity, plot (c) displays the information of plot (b) in the region between 40 GeV and 80 GeV.
The distributions in blue and red (black) color are the results obtained using the fixed scale $Q=M_Z$ and the running scale 
$Q=M_{\ell^+ \ell^-} (p_T(\ell^-))$ for $\alpha_s$, respectively. The dotted straight lines mark the results at the inclusive level.}
\label{fig:dist1}
\end{figure} 
For the fixed $\alpha_s$, the mixed QCD-electroweak corrections relative to the LO are roughly the same in all the bins and are of the order of 0.27\%, as seen in Fig.~\ref{fig:dist1}(a). This suggests that the nature of the events for the LO and two-loop corrections is kinematically similar. The two-loop corrections for the running $\alpha_s$ differ in each bin and are higher in the lower mass bins. 
It can go beyond 0.35\% in the lower bins, depending on the choice of the running scale. For the invariant mass above the $Z$ pole, the corrections for the running $\alpha_s$ with $Q=M_{\ell^+ \ell^-}$ become slightly smaller than the corrections for the fixed $\alpha_s$. This behavior is dictated by the one-loop running of $\alpha_s$.

\begin{figure}[ht!]
\centering
\subfloat[]{\includegraphics[width=0.6\linewidth]{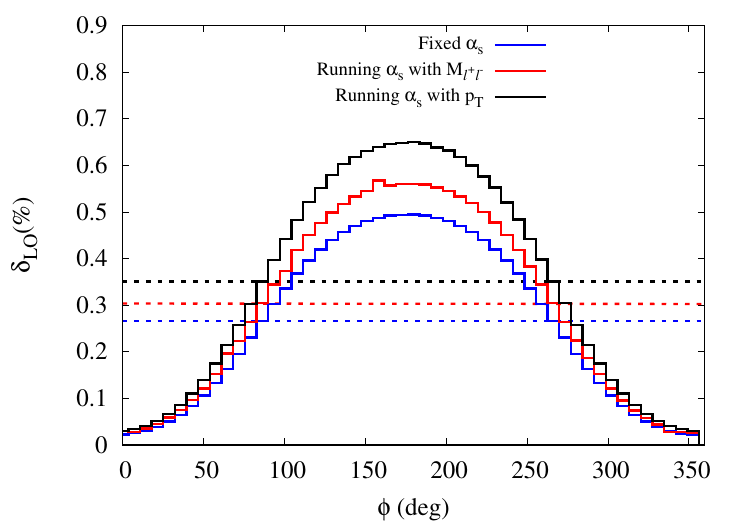}}
\newline
\centering
\subfloat[]{\includegraphics[width=0.5\linewidth]{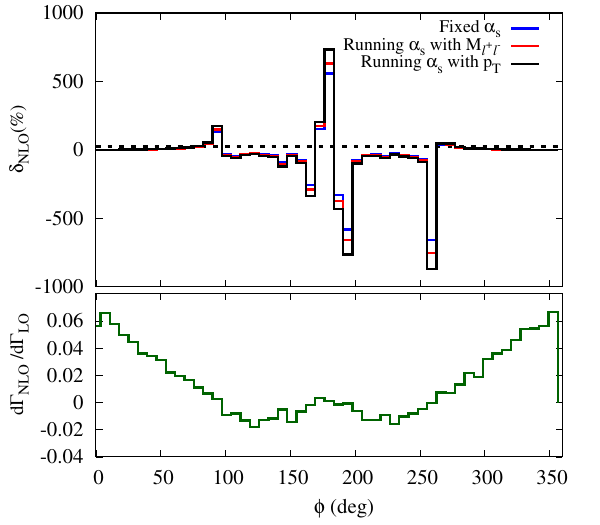}}
\subfloat[]{\includegraphics[width=0.5\linewidth]{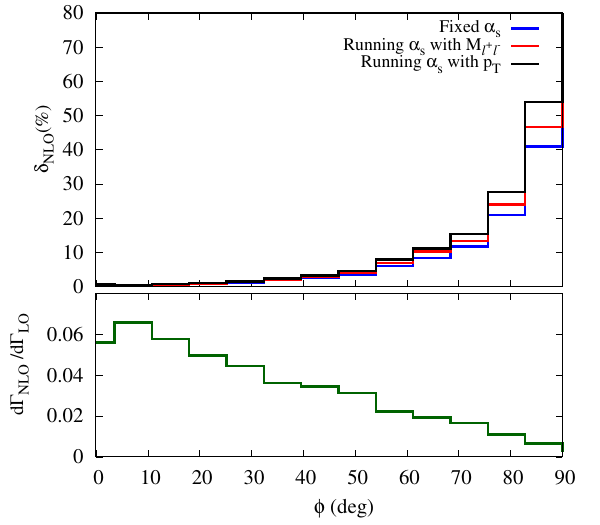}}
\caption{Effect of the mixed QCD-electroweak corrections on the distribution for the angle $\phi$ between the decay planes of the intermediate $Z$-bosons.
In (a), $\delta_{\rm LO}$ is the ratio of the mixed QCD-electroweak correction and the LO contribution. In the upper panels of (b) and (c), $\delta_{\rm NLO}$ is the ratio of the mixed QCD-electroweak correction and the NLO EW correction. The quantity $\rm {d\Gamma_{\rm NLO}/d\Gamma_{\rm LO}}$ in the lower panels of (b) and (c) gives the ratio of the NLO EW correction and the LO contribution.
For clarity, plot (c) displays the information of plot (b) in the region between 0 and $\dfrac{\pi}{2}$.
The distributions in blue and red (black) color are the results obtained using the fixed scale $Q=M_Z$ and the running scale 
$Q=M_{\ell^+ \ell^-} (p_T(\ell^-))$ for $\alpha_s$, respectively. The dotted straight lines mark the results at the inclusive level.}
\label{fig:dist3}
\end{figure}

In Fig.~\ref{fig:dist1}(b), we plot the two-loop corrections with respect to the NLO EW corrections in the upper panel. Since the electroweak corrections are also sensitive to the kinematics of the events, we note that in certain bins, between 20 GeV and 40 GeV, $\delta_{\rm NLO}$ becomes quite large independent of the choice of the scale $Q$. 
 {The relative effect of the NLO EW corrections with respect to the LO contributions in each bin, as shown in the lower panel, can be referred to understand the features of the distribution in the upper panel of Fig.~\ref{fig:dist1}(b). Significantly large values of $\delta_{\rm NLO}$ in certain bins are simply due to the fact that the NLO EW corrections are negligible in those bins. 
 This observation can be useful for the bin-wise analysis of the data. The two-loop corrections appear flat in the bins between 40 GeV and 80 GeV. However, a closer look reveals that $\delta_{\rm NLO}$ decreases in higher mass bins due to the larger NLO EW corrections in those bins. We have shown 
this in Fig.~\ref{fig:dist1}(c). The inverse behavior of the distributions in the upper and lower panels can be attributed to the kinematic similarity between the two-loop events and the LO events noted earlier.}

Apart from the invariant mass distribution, angular distributions are also effective in studying the Higgs properties. Therefore, we study the effect of the mixed QCD-electroweak corrections in the $\phi$ distribution, which is one of the most sensitive observables for the   
BSM studies. It is defined as the angle between the decay planes of the intermediate $Z$-bosons in the rest frame of the Higgs boson. This angle $\phi$ is the main observable for spin-parity assignment of the Higgs boson \cite{Soni:1993jc,Chang:1993jy,Skjold:1993jd,Barger:1993wt,Arens:1994wd,Buszello:2002uu,Choi:2002jk}. 
We have plotted the $\phi$ dependence of the relative corrections $\delta_i$ in Fig.~\ref{fig:dist3}.

In Fig.~\ref{fig:dist3}(a), we see that the mixed QCD-electroweak corrections relative to the LO do not exhibit a flat behavior in the $\phi$ distribution for the fixed $\alpha_s$. This is in contrast to what we see in the case of the invariant mass distribution in Fig.~\ref{fig:dist1}(a). We observe a $(1-\cos \phi)$ dependence in the shape of the $\phi$-distribution due to the mixed QCD-electroweak corrections. Also, the shape is independent of the scale choice for $\alpha_s$. 
The stated dependence is different from the LO behavior, which follows a {$\cos^2 \phi$} dependence~\cite{PhysRevD.37.1220,Choi:2002jk}. 
The difference can be attributed to the change in the effective $HZZ$ coupling due to the two-loop corrections. The two-loop corrections with respect to the LO are insignificant at the edges and large in the central region. The relative correction $\delta_{\rm LO}$ peaks at $\phi = \pi$. It is 0.49\% for the fixed and 0.56\% (0.64\%) for the running QCD coupling with $Q= M_{\ell^+ \ell^-}(p_T(\ell^-))$. Compared to the fixed $\alpha_s$, relative corrections are higher for the running $\alpha_s$ across all the bins. 

In the upper panel of Fig.~\ref{fig:dist3}(b), we have shown the relative effect of the two-loop corrections with respect to the NLO EW corrections. In the lower panel of the figure, the NLO EW corrections with respect to the LO are displayed. In the mid-region 
of the distribution, where the NLO EW corrections are negligible, more prominent peaks for the two-loop corrections can be seen. The numerical values of these peaks should not be taken very seriously. It is just that the two-loop corrections are more relevant in the bins with peaks, and they should be taken into account in the bin-wise analysis of the data aimed at BSM searches. $\delta_{\rm NLO}$ for $\phi$ looks flat near the edges. However, it is indeed not the case, as shown in Fig.~\ref{fig:dist3}(c) for the bins between 0 and $\frac{\pi}{2}$. In this range, $\delta_{\rm NLO}$ rises as $\rm {d\Gamma_{\rm NLO}/d\Gamma_{\rm LO}}$ goes down with increasing $\phi$. Similarly, in the region beyond $250^\circ$, as $\rm {d\Gamma_{\rm NLO}/d\Gamma_{\rm LO}}$ increases, $\delta_{\rm NLO}$ decreases (not shown explicitly) with an increasing value of $\phi$. These features are independent of the scale choice for $\alpha_s$. 

\section{Conclusions and outlook} 
\label{sec:conclusions}
In this paper, we have computed the mixed QCD-electroweak corrections to the partial decay width of the $H \to e^+ e^- \mu^+ \mu^-$ channel. It is one of the most crucial decay channels to study the Higgs properties at the LHC and for new physics searches in the Higgs sector. In-house codes are developed to systematically compute the contributing two-loop matrix elements at $\mathcal{O}(\alpha \alpha_s)$ using the projector technique. The bare two-loop matrix elements are found to be free from IR divergences but contain UV divergences, which are regularized using dimensional regularization and eliminated by the on-shell counter-term renormalization procedure. The whole calculation is implemented in the publicly available event generator {\tt{Hto4l}} code to perform the phase-space integration over the final state leptons and to obtain the improved predictions for the partial decay width. In the $G_\mu$ scheme, the mixed QCD-electroweak correction to the partial decay width for the Higgs mass of 125 GeV is found to be 0.27$\%$ (18\%)
of the LO prediction (NLO EW correction) for the fixed QCD coupling. With the running coupling, the corrections become larger. 
The corrections can be significantly larger for the Higgs mass above the $t \bar t$ threshold, which is relevant to various BSM scenarios. 

For the invariant mass distribution of the final state lepton pairs, the two-loop corrections with respect to the LO can cross 0.35\% in the lower bins depending on the scale choice.  
For the angular distributions, which are crucial in measuring the spin-$CP$ properties of the Higgs boson, the corrections with respect to the LO are of the order 0.6\% in the bins around $\phi =180^\circ$ when using the running $\alpha_s$. We also note that there are kinematic and angular bins in which the mixed QCD-electroweak corrections dominate the NLO EW corrections.
These results may be helpful in probing new physics in the Higgs sector. Needless to say, our computational framework also allows predictions for $H \to \gamma \gamma, \; \gamma Z, \;\gamma \ell^+ \ell^-, Z\ell^+ \ell^-$ decays. In addition to that, we can also use the ingredients calculated in this paper to predict $\mathcal{O}(\alpha \alpha_s)$ corrections for the $H \to \ell^+ \ell^- \ell^+ \ell^- \; (\ell = e,\mu)$.

\section*{Acknowledgement}
MK would like to acknowledge financial support from IISER Mohali for this work. MK would like to thank Biswajit Das for the useful discussions. XZ is supported by the Italian Ministry of Research (MUR) under grant PRIN 20172LNEEZ.
\appendix
\section{Projection operators}
\label{app:A}
The projectors required to obtain the form-factors given in Eq.~\ref{eqn:general amplitude} in $d=4-2\epsilon$ dimensions are given by, 
\begin{align}
P^A_{\mu\nu} &= \frac{1}{d-2}\left(g_{\mu\nu} + \frac{p_2^2 p_{1\mu}p_{1\nu}+ p_1^2 p_{2\mu}p_{2\nu} - (p_1.p_2) (p_{1\nu}p_{2\mu} + p_{1\mu}p_{2\nu})}{(p_1.p_2)^2 - p_1^2p_2^2}\right)\nonumber ,\\
P^B_{\mu\nu} &= \frac{1}{(d-2)((p_1.p_2)^2 - p_1^2p_2^2)^2}\bigg((p_1.p_2)^3\left(\frac{p_1^2p_2^2}{(p_1.p_2)^2}-1\right)g_{\mu\nu} +(d-1)(p_1.p_2)^2p_{1\mu}p_{2\nu}\nonumber\\
&-(d-1)(p_1.p_2)(p_2^2 p_{1\mu}p_{1\nu}+ p_1^2 p_{2\mu}p_{2\nu})+\left((p_1.p_2)^2+(d-2)p_1^2p_2^2\right)p_{1\nu}p_{2\mu}\bigg)\nonumber ,\\
P^C_{\mu\nu} &= \frac{\epsilon_{\mu\nu p_1p_2}}{(d-2)(d-3)((p_1.p_2)^2 - p_1^2p_2^2)}\nonumber ,\\
	P^D_{\mu\nu} &= \frac{1}{(d-2)((p_1.p_2)^2 - p_1^2p_2^2)^2}\bigg((p_1.p_2)^2 p_2^2\left(1-\frac{p_1^2p_2^2}{(p_1.p_2)^2}\right)g_{\mu\nu} + (p_2.p_2)^2(d-1)p_{1\mu}p_{1\nu}\nonumber\\
&-(d-1)p_2^2(p_1.p_2)(p_{1\mu}p_{2\nu} + p_{1\nu}p_{2\mu})+ (p_1^2p_2^2+(d-2)(p_1.p_2)^2)p_{2\mu}p_{2\nu})\bigg)\nonumber ,\\
	P^E_{\mu\nu} &= \frac{1}{(d-2)((p_1.p_2)^2 - p_1^2p_2^2)^2}\bigg((p_1.p_2)^2 p_1^2\left(1-\frac{p_1^2p_2^2}{(p_1.p_2)^2}\right)g_{\mu\nu} + (p_1.p_1)^2(d-1)p_{2\mu}p_{2\nu}\nonumber\\
&-(d-1)p_1^2(p_1.p_2)(p_{1\mu}p_{2\nu} + p_{1\nu}p_{2\mu})+ (p_1^2p_2^2+(d-2)(p_1.p_2)^2)p_{1\mu}p_{1\nu})\bigg)\nonumber ,\\
P^F_{\mu\nu} &= \frac{1}{(d-2)((p_1.p_2)^2 - p_1^2p_2^2)^2}\bigg((p_1.p_2)^3\left(\frac{p_1^2p_2^2}{(p_1.p_2)^2}-1\right)g_{\mu\nu} +(d-1)(p_1.p_2)^2p_{2\mu}p_{1\nu}\nonumber\\
&-(d-1)(p_1.p_2)(p_1^2 p_{2\mu}p_{2\nu}+ p_2^2 p_{1\mu}p_{1\nu})+\left((p_1.p_2)^2+(d-2)p_1^2p_2^2\right)p_{2\nu}p_{1\mu}\bigg)\nonumber .\\
\end{align}
\section{Transformation of parameters from the on-shell scheme to the complex-mass scheme}
\label{app:B}
The relation between the two sets of values $(M_V^{\rm OS},\Gamma_V^{\rm OS})$ and $(M_V,\Gamma_V)$ is given by~\cite{LHCHiggsCrossSectionWorkingGroup:2016ypw,Denner_2020}
\begin{equation}
M_V = \frac{M_V^{\rm OS}}{\sqrt{1+\bigg(\frac{\Gamma_V^{\rm OS}}{M_V^{\rm OS}}\bigg)^2}},\;\; \;\;\Gamma_V = \frac{\Gamma_V^{\rm OS}}{\sqrt{1+\bigg(\frac{\Gamma_V^{\rm OS}}{M_V^{\rm OS}}\bigg)^2}} \; \; \; \;\; (V=W,Z),
\end{equation}
resulting in
\begin{align}
   M_Z  &= 91.1535 \;{\rm GeV},& M_W  &= 80.3505 \;{\rm GeV}, \\
\Gamma_Z &= 2.49427 \;{\rm GeV},& \Gamma_W&= 2.1402 \;{\rm GeV}.
\end{align}
\newpage
\bibliography{literature} 
\bibliographystyle{JHEP}
\end{document}